\documentclass[aps,twocolumn, amsmath,amssymb,notitlepage,superscriptaddress,10pt]{revtex4-2}
\usepackage[utf8]{inputenc}
\usepackage{lineno,hyperref}
\usepackage{pifont}
\usepackage{amsmath}
\usepackage{textcomp}
\usepackage{natbib}
\usepackage{fleqn}
\usepackage[english]{babel}
\usepackage[autostyle]{csquotes}
\usepackage{graphicx}
\usepackage{txfonts}
\usepackage{tabularx}
\usepackage{flushend}
\usepackage{adjustbox}
\usepackage{xcolor}
\usepackage{booktabs}
\usepackage{multirow}
\usepackage{graphicx}
\usepackage{dcolumn}
\usepackage{bm}

\begin{document}
	
	
	\title{Ab-initio study of stable \textit{3d}, \textit{4d} and \textit{5d} transition metal  based Quaternary Heusler compounds}
	
	\author{S.~Nepal}
	\email{sashinpl@udel.edu}
	\affiliation{Department of Physics and Astronomy, University of Delaware, Newark, Delaware, 19716, United States of America.}
	\affiliation{Central Department of Physics, Tribhuvan University, Kathmandu, Nepal.}
	
	\author{R.~Dhakal}%
	\email{dhakr21@wfu.edu}
	\affiliation{Department of Physics, Wake Forest University, Winston-Salem, NC 27109, United States of America.}%
	\affiliation{Central Department of Physics, Tribhuvan University, Kathmandu, Nepal.}
	
	\author{I.~Galanakis}
	\email{galanakis@upatras.gr}%
	\affiliation{Department of Materials Science, School of Natural Sciences, University of Patras, GR-26504 Patras, Greece}%
	
	\author{S.M. Winter}
	\email{winters@wfu.edu}
	\affiliation{Department of Physics, Wake Forest University, Winston-Salem, NC 27109, United States of America.}	
	
	\author{R.~P.~Adhikari}
	\email{rajendra.adhikari@ku.edu.np}
	\affiliation{Department of Physics, Kathmandu University, Dhulikhel, Nepal.}%
	
	\author{G.~C.~Kaphle}
	\email{gopi.kaphle@cdp.tu.edu.np}
	\affiliation{Central Department of Physics, Tribhuvan University, Kathmandu, Nepal.}
	
	\date{\today}
	
\begin{abstract}
	The realization of the stable structure of Heusler compounds and the study of different properties is an important step for their potential application in spintronics and magnetoelectronic devices. In this paper, using the plane-wave pseudopotential method within the framework of density functional theory  (DFT), we investigate 25 Quaternary Heusler compounds for their electronic, magnetic, and mechanical properties. The Open Quantum Materials Database (OQMD) is used to screen a large number of compounds to narrow down the possible synthesizable materials. The convex-hull distance and elastic constants are exploited to confirm the thermodynamic and mechanical stability of the compounds. The careful study of the different structures suggests that 5 of the compounds crystallize in type-1 structure whereas 20 compounds adopt type-3 structure. The possible explanation for the observed behavior is made by invoking electronegativity arguments and through the study of individual spin magnetic moments in different structures. The compounds with diverse electronic and magnetic properties such as half-metallicity, spin gapless semiconducting behavior, and non-magnetic semi-conducting property have been identified.                
\end{abstract}

\maketitle

\emergencystretch 3em

\section{Introduction}
Since the prediction of half-metallicity in NiMnSb by de Groot \textit{et al.}\cite{deGroot} in 1983, the study of Heusler compounds has become an active research area as this particular property could be used in spintronics devices to increase their efficiency. Half-metals are compounds in which the electrons in two different spin channels show entirely contrasting characteristics allowing the complete spin polarization around the Fermi level. They are promising candidates for spintronics since both spin and charge can be exploited to manipulate information storage capacity, volatility and size and shape of the devices. The unprecedented development in the field of technology and science has enabled us to design new materials with different functionalities but at the same time, their application in real spintronics devices is marred by various challenges. The ideal candidate for spintronics materials should have qualities like low defects and disorder, high magnetoresistance, stability against thermal fluctuation and a  Curie temperature higher than room temperature \cite{spintronics, Hirohata,SpintronicsClaudia,HMAlloys-Galanakis-lec}. Furthermore, there should be similarity in crystal structure and lattice constant of the proposed half-metallic compounds with semi-conductors so that epitaxial growth is viable and hence one can achieve highly polarized films necessary for spin injection into a semi-conductor. These stringent requirements demand rigorous and careful theoretical as well as an experimental investigation of different potential candidates to systematically engineer these compounds for various applications.  
 
\par Besides Heusler alloys, the half-metallic property has been reported in different classes of magnetic materials such as double-perovskites\cite{kato2004structural}, diluted magnetic semiconductors\cite{stroppa2003electronic,akai1998ferromagnetism}, manganites\cite{soulen1998measuring} and number of Zinc-blende compounds of  the transition metal elements with the \textit{sp} elements\cite{galanakis2003zinc}. However, in the last two decades, Heusler alloys have become increasingly popular due to the development of general structure-property relations, which make it possible to anticipate electronic and magnetic properties\cite{Galanakis2002,Galanakisinverse,ozdougan2013slater}. Moreover, the high Curie temperature of Heusler compounds\cite{wurmehl2005geometric} make them appropriate candidates for room temperature applications whereas their compatible crystal structure with semi-conductors render them suitable for nearly perfect injection of spin-polarized current into semiconductors.

\par Heusler compounds are extensively studied ternary intermetallics with a large number of members having vast varieties of magnetic properties\cite{HeuslerPropandGrowth, fong2013half}. Many members of the Heusler family are reported to have diverse phenomena extending from magnetic semiconductors, spin-gapless semiconductors, half-metals to topological insulators. The general formula to represent Heusler compounds is X$_2$YZ, XYZ or XX\textquotesingle YZ, where X, X\textquotesingle, and Y are transition metal atoms and Z is a main-group element.  Depending on the configuration and number of elements involved, the compounds can be full, inverse, semi or quaternary Heusler alloys. All types of Heusler compounds crystallize with four interpenetrating face-centered cubic (fcc) structures. In the case of semi-Heusler compounds, among four sublattices, one is unoccupied whereas in other structures all are fully occupied. In our discussion, these four sublattices are denoted by A, B, C, and D with their respective Wyckoff position as (0,0,0), ($\frac{1}{4}$,$\frac{1}{4}$,$\frac{1}{4}$) ($\frac{1}{2}$,$\frac{1}{2}$,$\frac{1}{2}$) and ($\frac{3}{4}$,$\frac{3}{4}$,$\frac{3}{4}$).
 
\par The relative positions of each atom are often dictated by the electronegativity argument\cite{SpintronicsClaudia}. Since the elements in the Heusler compounds are covalently bonded, atoms with a small difference in electronegativity are preferred in the different sublattices during the formation of the compounds. In full Heusler compounds, since the valence of X atom is greater than that of Y, the two X atoms occupy A and C coordinates whereas Y atoms positioned themselves in B coordinates and hence the arrangement of the atoms is X-Y-X-Z along the diagonal. Among the two coordinates A and C, when one position is unoccupied, the structure reduces to  semi or half-Heusler compounds. The inverse Heusler alloy distinguishes itself from  the full Heusler structure in terms of the position of X and Y atoms in the sublattices; here the valence of Y atoms is greater than that of X due to which two X atoms take up A and B coordinates resulting in the occupation of C coordinates by Y atoms. In the unit cell, the sequence of diagonal elements in this case looks like X-X-Y-Z. When we replace one of the two X atoms in full Heusler alloys by a new transition metal atoms X\textquotesingle$\,$   , the structure transforms to quaternary Heusler alloys. It is important to note that though the generic formula of quaternary Heusler compounds is XX\textquotesingle$\,$YZ, the sequence of atoms along the diagonal of a unit cell is X-Y-X\textquotesingle$\,$-Z. For the structure of quaternary Heusler compounds, three different possible  non-equivalent configurations exist which would be discussed in detail in the next section. 
\begin{figure}[h]
	\centering
	\includegraphics[width=0.7\linewidth]{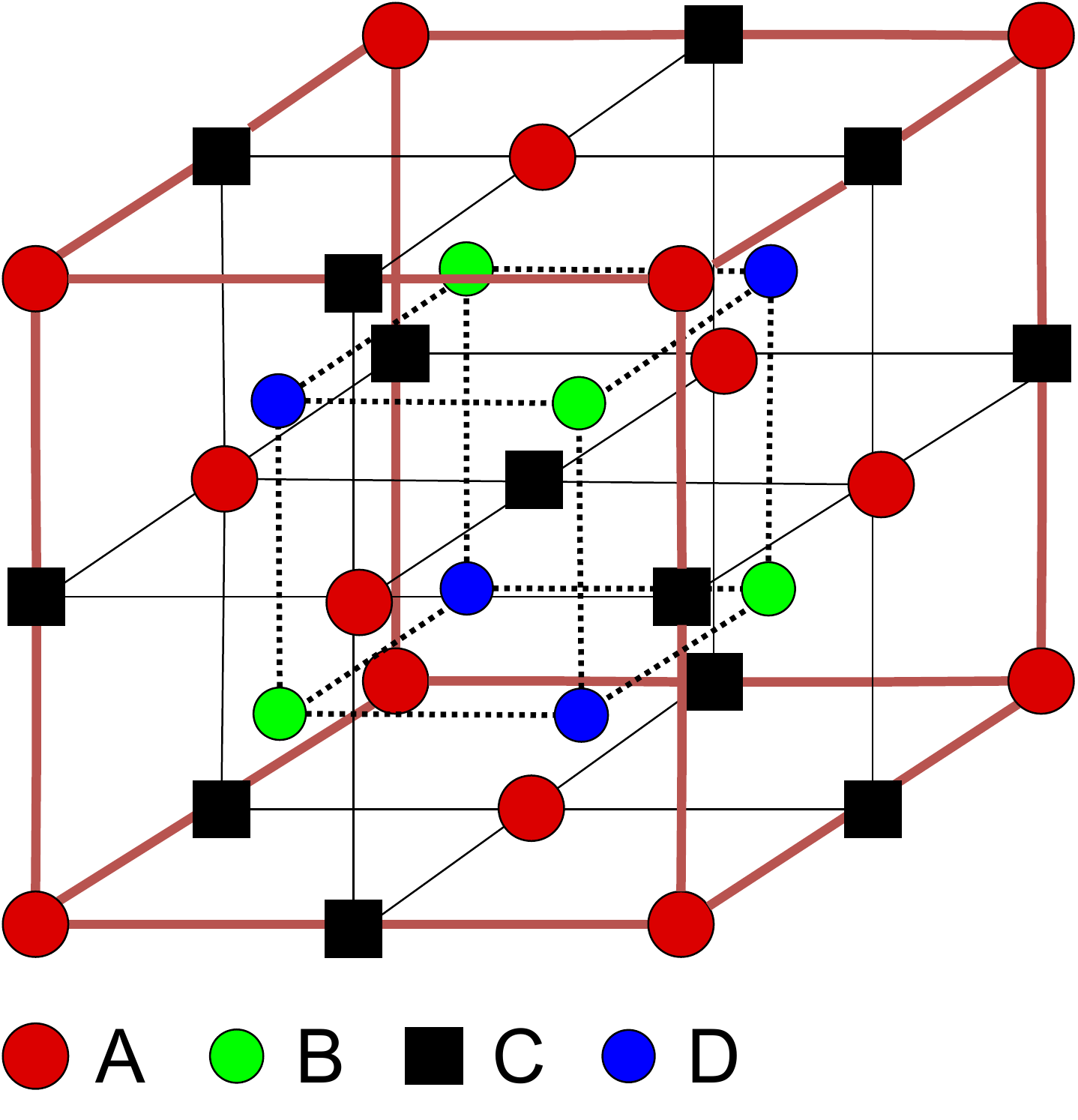}
	\caption{Schematic representation of the general cubic lattice structure of Heusler compounds consisting of four sites per unit cell. Various types of Heusler compounds can be formed by intermixing different atoms on  the respective crystallographic coordinates, leading to different symmetries and structures as explained in the text.}
	\label{fig:heusler}
\end{figure}

\par The possible application of quaternary Heusler alloys in spintronics was at first pointed out by Block \textit{et al.}\cite{block2003large, felser2003investigation} who reported a huge negative magnetoresistivity in Co$_2$Cr$_{0.6}$Fe$_{0.4}$Al  at room temperature in  the presence of a small external field. The next year, Galanakis predicted half-metallicity in several quaternary Heusler compounds\cite{galanakis2004appearance} by using first-principles calculations. The theoretical and experimental investigation of CoFeMnSi\cite{dai2009new} by Dai \textit{et al.}, in 2009, demonstrating a large half-metallic gap triggered the study of quaternary Heusler compounds with a 1:1:1:1 stoichiometry. Alijani \textit{et al.}, a few years later, extended this work to CoFeMnZ(Z = Al, Ga, Si, Ge) compounds\cite{alijani2011electronic} reporting all of them as half-metallic ferromagnets. In another separate work, for the first time, they synthesized Ni-based quaternary Heusler compounds and predicted half-metallicity in those compounds\cite{alijani2011quaternary}. In 2013 Gao and collaborators, by using the ab initio method, studied CoFeCr-based series\cite{gao2013large} and reported a large half-metallic gap in CoFeCrSi which was robust against the lattice compression and inclusion of on-site electrostatic Coulomb interaction. Recently, Gao and their team has studied a large number of quaternary Heusler compounds by using high-throughput density functional theory (HT DFT) screening method and identified 70 compounds as stable spin gapless semiconductors(SGSs)\cite{gao2019high}. The dataset was exploited by Aull \textit{et al.}\cite{aull2019ab} to  identify type-I and type-II SGSs with large gaps, and used it to predict the potential candidates for reconfigurable magnetic tunnel diodes and transistors. 
       
\section{Motivation and Aim}

\par In this paper, motivated by the above-mentioned work, we aim to screen a large number of quaternary Heusler compounds in the search of materials with novel properties like half-metallicity, spin gapless semiconducting behavior, etc. In order to achieve this goal we exploit the Open Quantum Materials Database (OQMD)\cite{OQMD1,OQMD2}. This database has been quite successful for accurately predicting the elemental groundstate structures of the compounds\cite{OQMD2}. In our previous work, we have used OQMD to investigate several FeCr-based quaternary Heusler alloys with interesting properties\cite{dhakal2021prediction}. The scope of this paper is to extend our previous work more systematically to a large number of quaternary Heusler alloys and pave a way for experimentalist in this research area to grow these compounds with predetermined electronic and magnetic properties.

\par In this work, we have investigated  quaternary Heusler alloys where the constituent elements in the compounds are from 3\textit{d}, 4\textit{d} and 5\textit{d} series of transition metal atoms. In the literature discussed above, the majority of quaternary Heusler alloys have magnetic atoms with partially filled 3\textit{d} shells. Hence, it would be edifying and interesting to analyze and compare the results of the compounds containing only 3\textit{d} electrons with the compounds containing the mixture of 3\textit{d}, 4\textit{d} and 5\textit{d}  elements. It has been observed experimentally that the substitution of 3\textit{d} elements in equiatomic quaternary Heusler alloys with 4\textit{d} elements improves the degree of disorder and increases the Curie temperature\cite{bainsla2015corufex}. Furthermore, the half-metallicity of such compounds are robust against lattice parameter variation and tetragonal distortion\cite{wang2017structural}.  In addition, due to strong hybridization between 3\textit{d} and 4\textit{d} or 5\textit{d} valence states, one can expect a large gap in quaternary Heusler alloys whose constituent elements are  4\textit{d} or 5\textit{d} atoms.  A few attempts have been made to study such compounds theoretically
 \cite{berri2014robust,kundu2017new,labar2021novel,wang2017first,forozani2020structural,chinnadurai2021first,pughcriticalelastic_Seh} but the initial scanning of Convex-hull distance of the majority of compounds mentioned in the literature shows a value greater than 0.20 eV/atom making them difficult to synthesize experimentally.
 \par The realization of  stable structures  that are experimentally feasible among different possible configurations is one of the major challenges of computational materials science. Hence, it is very important that we apply different chemical and energy constraints to rule out the compounds which are unlikely to be formed.  Among different thermodynamic constraints, the formation energy is frequently used to predict the stability of the compound. It can be defined as the energy difference  between the total energy of the bulk compound and the sum of the energies of the constituent atoms in the elemental phase. The negative value of formation energy is necessary to grow a compound in a given structure. However, it is not sufficient to predict whether a particular structure is stable against other similar phases at the given stoichiometry. The convex-hull, which for a given stoichiometry can be defined as the phase with the minimum energy among different studied phase, is an energy quantity that can be used to more reliably predict whether the given structure could be realized experimentally.
 \par In this communication, we have used OQMD to filter out the compounds by setting proper threshold for convex-hull distance. From the large pool of sample structures, we first make a list of promising quaternary Heusler compounds whose convex-hull distance is less than 0.2 eV/atom. We have chosen this value because we believe almost all (meta)stable phases can be found within this distance from the convex-hull. The justification to set this particular threshold for convex-hull is described in detail in our previous work (see reference \cite{dhakal2021prediction}). We have investigated the quaternary Heusler compounds where the elements from 4\textit{d} and 5\textit{d}  series like Zr, Ru, Rh and Ir combine with the elements from 3\textit{d} series like Cr, Mn, Fe, and Co, resulting in compounds with interesting electronic and magnetic properties. Out of twenty-five  compounds we study, we find seven compounds with large half-metallic gap, fourteen are nearly half-metals, two of the compounds shows spin gapless semiconducting behavior and two are non-magnetic semiconductors. To the best of our knowledge, six of the compounds have been reported earlier\cite{ozdougan2013slater,recent1,recent2, recent3, recent4, recent5, recent6, recent7}. Among these six compounds, CoCrZrAl, CoRuCrSi, and RhFeTisi are half-metals where as CoRuCrGe, FeRuCrGe, CoMnCrSi are nearly half-metals. It is important to note that CoMnCrSi was initially assumed to be crystallized in type-3 structure\cite{recent7,ozdougan2013slater} as it is the preferred structure in most of the quaternary Heusler compounds but our careful calcualtion shows that CoMnCrSi along with other four compounds crystallize in type-1 structure. Attempts are made to understand this peculiar behavior.

\section{Computational details and methodology}
Ab initio calculations are carried out by exploiting plane-wave pseudopotential method as provided by Quantum Espresso package\cite{QE-2009,QE-2017} within the framework of density functional theory (DFT). We have used ultrasoft pseudopotentials\cite{ultrasoft} for all the elements. The exchange-correlation potential is estimated by generalized gradient approximation of Perdew-Burke-Ernzerhof (PBE-GGA)\cite{perdew1996generalized} while calculating the structural, electronic, and magnetic properties of the compounds. The cut-off energy for plane wave expansion is set to 100 $Ry$ with a Monkhorst-Pack grid of $8\times8\times8$ during the self-consistent field (SCF) calculation. A denser k-mesh of $16\times16\times16$ is used to calculate the density of states. The threshold for the convergence of force, total energy, and scf cycle is set to $10^{-4} Ry/a.u$, $10^{-5}Ry$, and $10^{-8}$ respectively. The total energy minimization method is used for structural optimization and a linear tetrahedron scheme\cite{lineartetrahedra} is implemented for Brillouin zone integration.
\par We start by collecting a large number of quaternary Heusler compounds by combining different elements from the periodic table. Compounds thus collected are checked in OQMD for the minimum threshold convex-hull energy and negative formation energy. We discard the compounds which do not satisfy the mentioned criteria. If the compound has a convex hull distance less than 0.2eV/atom and negative formation energy, we investigate the compound in different magnetic configurations and three nonequivalent structures by using the lattice parameter provided by the OQMD.
\begin{figure}[h!]
	\centering 
	\includegraphics[width=1\linewidth]{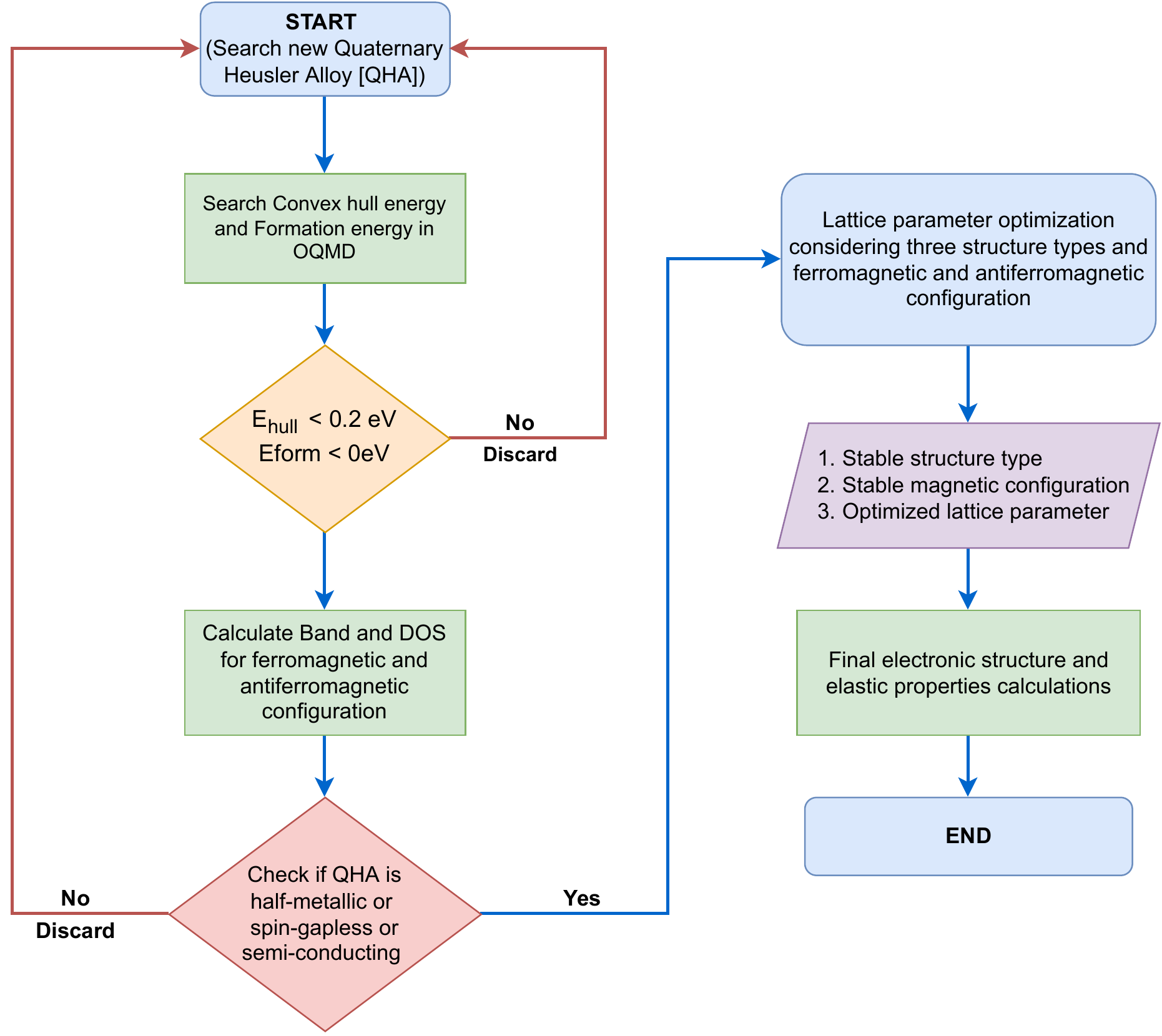}
	\caption{The flowchart for the algorithm describing the process for the identification and selection of the Heusler compounds studied in this paper.}
	\label{fig:flowchart}
\end{figure}

 After careful analysis of the total density of states, if we find the compound to be promising, we start with lattice parameter optimization (see Figure \ref{fig:flowchart}). This optimized lattice parameter is used to determine the stable structure by considering different magnetic configurations and nonequivalent structures in order to calculate structural, electronic, and magnetic properties of the compound.
 
 \section{Result and discussions}
 
 \subsection{Ground state Structures}
 The quaternary Heusler compounds, whose prototype material is LiMgPdSb, crystallize in the so-called Y-structure with space group F$\bar{4}$3m(216).  In principle, they can crystallize in three different non-equivalent superstructure\cite{dai2009new,alijani2011electronic} as shown in Table \ref{tab:table-I}. Unlike metals, due to the possibility of the existence of multiple local minima, the ground state calculation of magnetic materials is more complex. Hence, we assume the ferromagnetic as well as anti-ferromagnetic configurations to determine the ground state of the studied compounds. The total energies of both magnetic configurations are compared to find out the stable structure in the given stoichiometry (not shown in Table \ref{tab:my-table}). The comparison of total energies, optimized lattice parameter, and total spin magnetic moment of the compounds in three different non-equivalent superstructures is shown in Table \ref{tab:my-table}.  The optimized structure with minimum energy in a stable initial magnetic configuration suggests that among 25 reported compounds, 5 of the compounds crystallize in type-1 structure whereas the rest of the compounds prefer to be stabilized in type-3 structure. The convergence is achieved nicely for all the three superstructures except FeRuCrGe for which we are not able to converge type-2 structure.

\begin{table}[h]
	\caption{\label{tab:table-I}Different possible position of X, X\textquotesingle$\,$, Y and Z atoms in three non-equivalent configurations of quaternary Heusler compounds. Here, X, X\textquotesingle$\,$     and Y are transition metal atoms whereas Z is a  main-group element.}
	
	\begin{ruledtabular}
		\begin{tabular}{lcccc}
			\multirow{2}{*}{} & A       & B             & C             & D             \\ \cline{2-5}
			& (0,0,0) & ($\frac{1}{4}$,$\frac{1}{4}$,$\frac{1}{4}$) & ($\frac{1}{2}$,$\frac{1}{2}$,$\frac{1}{2}$) & ($\frac{3}{4}$,$\frac{3}{4}$,$\frac{3}{4}$) \\ \bottomrule[0.05em]
			Type-1            & Z       & X            & X\textquotesingle$\,$               & Y             \\ 
			Type-2           & Z       & X\textquotesingle$\,$              & X            & Y            \\ 
			Type-3          & X      & Z             & X\textquotesingle$\,$               & Y             \\
		\end{tabular}%
	\end{ruledtabular}
\end{table}
\begin{table*}[t!]
\caption{Comparison of total energies (E$_{tot}$), optimized lattice parameters (a$_{opt}$) and magnetic moments (m$_{tot}$) of type-1, type-2 and type-3 structures. Here, N$_{v}$ represents the total number of valence electrons of a given compound.}
	\label{tab:my-table}
	\begin{ruledtabular}
		\begin{tabular}{lcccclcccclcclcll}
			\multirow{2}{*}{} & \multicolumn{1}{l}{\multirow{2}{*}{N$_V$}} & \multicolumn{1}{l}{\multirow{2}{*}{\begin{tabular}[c]{@{}l@{}}E$_{tot}$\\ (Rydberg)\end{tabular}}} & \multirow{2}{*}{\begin{tabular}[c]{@{}c@{}}a$_{opt}$\\ ($\AA$)\end{tabular}} & m$_{tot}$                         & \multirow{2}{*}{} & \multicolumn{1}{l}{\multirow{2}{*}{N$_V$}} & \multicolumn{1}{l}{\multirow{2}{*}{\begin{tabular}[c]{@{}l@{}}E$_{tot}$\\ (Rydberg)\end{tabular}}} & \multirow{2}{*}{\begin{tabular}[c]{@{}c@{}}a$_{opt}$\\ ($\AA$)\end{tabular}} & m$_{tot}$                      & \multirow{2}{*}{} & \multicolumn{1}{l}{\multirow{2}{*}{N$_V$}} & \multicolumn{1}{l}{\multirow{2}{*}{\begin{tabular}[c]{@{}l@{}}E$_{tot}$\\ (Rydberg)\end{tabular}}} & \multicolumn{1}{c}{\multirow{2}{*}{\begin{tabular}[c]{@{}c@{}}a$_{opt}$\\ ($\AA$)\end{tabular}}} & m$_{tot}$                                     &  &  \\ 
			& \multicolumn{1}{l}{}                       & \multicolumn{1}{l}{}                                                                               &                                                                                           & \textbf{($\mu_B$)} &                   & \multicolumn{1}{l}{}                       & \multicolumn{1}{l}{}                                                                               &                                                                                           & ($\mu_B$)       &                   & \multicolumn{1}{l}{}                       & \multicolumn{1}{l}{}                                                                               & \multicolumn{1}{c}{}                                                                                          & \multicolumn{1}{l}{($\mu_B$)} &  &  \\ \bottomrule[0.05em]
			\textbf{CoCrZrAl} &                                             &                                                                                                     &                                                                                           &                                   & \textbf{IrCoTiAl} & \textbf{}                                   & \textbf{}                                                                                           & \textbf{}                                                                                 & \textbf{}                      & \textbf{RhCrTiAl} & \multicolumn{1}{l}{\textbf{}}              & \multicolumn{1}{l}{\textbf{}}                                                                      & \textbf{}                                                                                                      & \multicolumn{1}{l}{\textbf{}}                &  &  \\ 
			Type-I            & \multirow{3}{*}{22}                         & -570.94643                                                                                          & 6.653                                                                                     & 5.35                              & Type-I            & \multirow{3}{*}{25}                         & -478.52686                                                                                          & 6.0621                                                                                    & 1.90                           & Type-I            & \multirow{3}{*}{22}                         & -526.32351                                                                                          & \multicolumn{1}{c}{6.1795}                                                                                    & 4.00                                          &  &  \\  
			Type-II           &                                             & -571.03889                                                                                          & 6.3065                                                                                    & 4.44                              & Type-II           &                                             & -478.49861                                                                                          & 6.0904                                                                                    & 2.48                           & Type-II           &                                             & -526.31386                                                                                          & \multicolumn{1}{c}{6.2315}                                                                                    & 3.98                                          &  &  \\  
			Type-III          &                                             & -571.08059                                                                                          & 6.2422                                                                                    & 4.00                              & Type-III          &                                             & -478.63703                                                                                          & 6.0212                                                                                    & 1.00                           & Type-III          &                                             & -526.39912                                                                                          & \multicolumn{1}{c}{6.0996}                                                                                    & -2.00                                          &  &  \\ 
			\textbf{CoIrMnSb} &                                             &                                                                                                     &                                                                                           &                                   & \textbf{IrFeZrAl} & \textbf{}                                   & \textbf{}                                                                                           & \textbf{}                                                                                 & \textbf{}                      & \textbf{RhFeMnGe} & \multicolumn{1}{l}{\textbf{}}              & \multicolumn{1}{l}{\textbf{}}                                                                      & \textbf{}                                                                                                      & \multicolumn{1}{l}{\textbf{}}                &  &  \\ 
			Type-I            & \multirow{3}{*}{30}                         & -587.28162                                                                                          & 6.1794                                                                                    & 4.67                              & Type-I            & \multirow{3}{*}{24}                         & -414.42794                                                                                          & 6.2838                                                                                    & 2.78                           & Type-I            & \multirow{3}{*}{28}                         & -699.97105                                                                                          & \multicolumn{1}{c}{5.9475}                                                                                    & 5.28                                          &  &  \\  
			Type-II           &                                             & -587.29265                                                                                          & 6.2033                                                                                    & 4.68                              & Type-II           &                                             & -414.39204                                                                                          & 6.3626                                                                                    & 3.89                           & Type-II           &                                             & -699.93485                                                                                          & \multicolumn{1}{c}{6.0225}                                                                                    & 0.52                                          &  &  \\  
			Type-III          &                                             & -587.35666                                                                                          & 6.2390                                                                                     & 6.00                              & Type-III          &                                             & -414.50656                                                                                          & 6.2182                                                                                    & 0.00                          & Type-III          &                                             & -699.98596                                                                                          & \multicolumn{1}{c}{5.9089}                                                                                    & 4.07                                          &  &  \\ 
			\textbf{CoIrMnSn} &                                             &                                                                                                     &                                                                                           &                                   & \textbf{IrMnCrGe} & \multicolumn{1}{l}{\textbf{}}              & \multicolumn{1}{l}{\textbf{}}                                                                      & \multicolumn{1}{l}{\textbf{}}                                                            & \multicolumn{1}{l}{\textbf{}} & \textbf{RhFeMnSi} & \multicolumn{1}{l}{\textbf{}}              & \multicolumn{1}{l}{\textbf{}}                                                                      & \textbf{}                                                                                                      & \textbf{}                                     &  &  \\ 
			Type-I            & \multirow{3}{*}{29}                         & -730.44900                                                                                            & 6.1999                                                                                    & 5.21                              & Type-I            & \multirow{3}{*}{26}                         & -462.71474                                                                                          & 6.0098                                                                                    & 2.00                          & Type-I            & \multirow{3}{*}{28}                         & -698.45909                                                                                          & \multicolumn{1}{c}{5.8088}                                                                                    & 4.75                                          &  &  \\  
			Type-II           &                                             & -730.46470                                                                                           & 6.2245                                                                                    & 5.31                              & Type-II           &                                             & -462.63227                                                                                          & 6.0425                                                                                    & 2.29                          & Type-II           &                                             & -698.40063                                                                                          & \multicolumn{1}{c}{5.8821}                                                                                    & 0.34                                          &  &  \\  
			Type-III          &                                             & -730.53115                                                                                          & 6.1999                                                                                    & 5.01                              & Type-III          &                                             & -462.68808                                                                                          & 5.9509                                                                                    & 2.01                           & Type-III          &                                             & -698.47764                                                                                          & \multicolumn{1}{c}{5.8021}                                                                                    & 4.01                                          &  &  \\ 
			\textbf{CoRuCrGa} &                                             &                                                                                                     &                                                                                           &                                   & \textbf{IrMnCrSi} & \multicolumn{1}{l}{\textbf{}}              & \multicolumn{1}{l}{\textbf{}}                                                                      & \multicolumn{1}{l}{\textbf{}}                                                            & \multicolumn{1}{l}{\textbf{}} & \textbf{RhFeTiGe} & \multicolumn{1}{l}{\textbf{}}              & \multicolumn{1}{l}{\textbf{}}                                                                      & \textbf{}                                                                                                      & \textbf{}                                     &  &  \\ 
			Type-I            & \multirow{3}{*}{26}                         & -834.69208                                                                                          & 5.8609                                                                                    & 1.83                              & Type-I            & \multirow{3}{*}{26}                         & -461.21042                                                                                          & 5.8909                                                                                    & 2.00                          & Type-I            & \multirow{3}{*}{25}                         & -605.94014                                                                                          & \multicolumn{1}{c}{6.1040}                                                                                     & 2.36                                          &  &  \\  
			Type-II           &                                             & -834.69461                                                                                          & 5.9831                                                                                    & 5.49                              & Type-II           &                                             & -461.10369                                                                                          & 5.9115                                                                                    & 2.43                          & Type-II           &                                             & -605.90347                                                                                          & \multicolumn{1}{c}{6.1409}                                                                                    & 3.41                                          &  &  \\  
			Type-III          &                                             & -834.75637                                                                                          & 5.8747                                                                                    & 2.05                              & Type-III          &                                             & -461.19109                                                                                          & 5.8485                                                                                    & 2.00                           & Type-III          &                                             & -605.97901                                                                                          & \multicolumn{1}{c}{6.0132}                                                                                    & 1.01                                          &  &  \\ 
			\textbf{CoRuCrGe} &                                             &                                                                                                     &                                                                                           &                                   & \textbf{IrRuTiAl} & \multicolumn{1}{l}{\textbf{}}              & \multicolumn{1}{l}{\textbf{}}                                                                      & \multicolumn{1}{l}{\textbf{}}                                                            & \multicolumn{1}{l}{\textbf{}} & \textbf{RhFeTiSi} & \multicolumn{1}{l}{\textbf{}}              & \multicolumn{1}{l}{\textbf{}}                                                                      & \textbf{}                                                                                                      & \multicolumn{1}{l}{\textbf{}}                &  &  \\ 
			Type-I            & \multirow{3}{*}{27}                         & -675.43013                                                                                          & 5.8971                                                                                    & 1.71                              & Type-I            & \multirow{3}{*}{24}                         & -384.01028                                                                                          & 6.2000                                                                                    & 0.28                           & Type-I            & \multirow{3}{*}{25}                         & -604.40378                                                                                          & \multicolumn{1}{c}{6.0115}                                                                                    & 2.36                                          &  &  \\  
			Type-II           &                                             & -675.43972                                                                                          & 5.8771                                                                                    & 2.48                              & Type-II           &                                             & -384.00804                                                                                          & 6.2016                                                                                    & 0.12                           & Type-II           &                                             & -604.35556                                                                                          & \multicolumn{1}{c}{6.0475}                                                                                    & 3.34                                          &  &  \\  
			Type-III          &                                             & -675.49712                                                                                          & 5.8992                                                                                    & 3.01                              & Type-III          &                                             & -384.21608                                                                                          & 6.1029                                                                                    & 0.00                            & Type-III          &                                             & -604.46226                                                                                          & \multicolumn{1}{c}{5.9147}                                                                                    & 1.00                                          &  &  \\ 
			\textbf{CoRuCrSi} & \textbf{}                                   & \textbf{}                                                                                           & \textbf{}                                                                                 & \textbf{}                         & \textbf{NiFeVAl}  & \multicolumn{1}{l}{\textbf{}}              & \multicolumn{1}{l}{\textbf{}}                                                                      & \multicolumn{1}{l}{\textbf{}}                                                            & \multicolumn{1}{l}{\textbf{}} & \textbf{RuCrTiSi} & \multicolumn{1}{l}{}                       & \multicolumn{1}{l}{}                                                                               &                                                                                                                &                                               &  &  \\ 
			Type-I            & \multirow{3}{*}{27}                         & -673.91813                                                                                          & 5.7911                                                                                    & 1.62                              & Type-I            & \multirow{3}{*}{26}                         & -738.09260                                                                                           & 5.8652                                                                                    & 1.80                           & Type-I            & \multirow{3}{*}{22}                         & \multicolumn{1}{l}{-502.14413}                                                                     & 5.9783                                                                                                         & 0.04                                          &  &  \\  
			Type-II           &                                             & -673.94635                                                                                          & 5.7519                                                                                    & 0.44                              & Type-II           &                                             & -738.04627                                                                                          & 5.8669                                                                                    & 2.03                           & Type-II           &                                             & \multicolumn{1}{l}{-502.11233}                                                                     & 6.0729                                                                                                         & 3.42                                          &  &  \\  
			Type-III          &                                             & -673.99479                                                                                          & 5.7930                                                                                     & 3.00                              & Type-III          &                                             & -738.12608                                                                                          & 5.7846                                                                                    & 1.99                           & Type-III          &                                             & \multicolumn{1}{l}{-502.21994}                                                                     & 5.9792                                                                                                         & -1.99                                          &  &  \\ 
			\textbf{CoRuZrSi} & \textbf{}                                   & \textbf{}                                                                                           & \textbf{}                                                                                 & \textbf{}                         & \textbf{NiMnCrAl} & \multicolumn{1}{l}{\textbf{}}              & \multicolumn{1}{l}{\textbf{}}                                                                      & \multicolumn{1}{l}{\textbf{}}                                                            & \multicolumn{1}{l}{\textbf{}} & \textbf{RuCrZrGa} & \multicolumn{1}{l}{}                       & \multicolumn{1}{l}{}                                                                               &                                                                                                                &                                               &  &  \\ 
			Type-I            & \multirow{3}{*}{25}                         & -598.33346                                                                                          & 6.2238                                                                                    & 2.91                              & Type-I            & \multirow{3}{*}{26}                         & -733.55295                                                                                          & 5.8250                                                                                     & 2.00                          & Type-I            & \multirow{3}{*}{21}                         & \multicolumn{1}{l}{-642.55333}                                                                     & 6.3413                                                                                                         & 2.97                                          &  &  \\  
			Type-II           &                                             & -598.39200                                                                                          & 6.1724                                                                                    & 2.17                              & Type-II           &                                             & \multicolumn{1}{l}{-733.48878}                                                                     & 5.9360                                                                                     & 0.27                          & Type-II           &                                             & -642.56131                                                                                          & 6.4093                                                                                                         & 3.43                                          &  &  \\  
			Type-III          &                                             & -598.47233                                                                                          & 6.1191                                                                                    & 1.00                              & Type-III          &                                             & \multicolumn{1}{l}{-733.54890}                                                                     & 5.7750                                                                                     & 1.88                           & Type-III          &                                             & -642.64791                                                                                          & 6.3195                                                                                                         & 3.00                                          &  &  \\ 
			\textbf{FeRuCrGe} & \textbf{}                                   & \textbf{}                                                                                           & \textbf{}                                                                                 & \textbf{}                         & \textbf{RhCoZrAl} & \multicolumn{1}{l}{\textbf{}}              & \multicolumn{1}{l}{\textbf{}}                                                                      & \multicolumn{1}{l}{\textbf{}}                                                            & \multicolumn{1}{l}{\textbf{}} & \textbf{RuMnCrSi} & \multicolumn{1}{l}{}                       & \multicolumn{1}{l}{}                                                                               &                                                                                                                &                                               &  &  \\
			Type-I            & \multirow{3}{*}{26}                         & -631.67285                                                                                          & 5.9227                                                                                    & 2.29                              & Type-I            & \multirow{3}{*}{25}                         & -622.59001                                                                                          & 6.2718                                                                                    & -2.08                          & Type-I            & \multirow{3}{*}{25}                         & \multicolumn{1}{l}{-596.26269}                                                                     & 5.8742                                                                                                         & 1.01                                         &  &  \\ 
			Type-II           &                                             & -                                                                                                   & -                                                                                         & -                                 & Type-II           &                                             & \multicolumn{1}{l}{-622.55551}                                                                     & 6.3092                                                                                    & 2.53                           & Type-II           &                                             & -596.15929                                                                                          & 5.8540                                                                                                         & 0.45                                          &  &  \\ 
			Type-III          &                                             & -631.73447                                                                                          & 5.8872                                                                                    & 2.01                              & Type-III          &                                             & \multicolumn{1}{l}{-622.70147}                                                                     & 6.2225                                                                                    & 0.91                           & Type-III          &                                             & -596.25793                                                                                          & 5.7837                                                                                                         & 1.01                                          &  & \\
			\textbf{CoMnCrSi} & \textbf{}                                   & \textbf{}                                                                                           & \textbf{}                                                                                 & \textbf{}                         & \textbf{} & \multicolumn{1}{l}{\textbf{}}              & \multicolumn{1}{l}{\textbf{}}                                                                      & \multicolumn{1}{l}{\textbf{}}                                                            & \multicolumn{1}{l}{\textbf{}} & \textbf{} & \multicolumn{1}{l}{}                       & \multicolumn{1}{l}{}                                                                               &                                                                                                                &                                               &  &  \\
			Type-I            & \multirow{3}{*}{26}                         & -690.73220                                                                                          & 5.6796                                                                                    & 2.01                             &             & \multirow{3}{*}{}                         &                                                                                          &                                                                                    &                          &             & \multirow{3}{*}{}                         & \multicolumn{1}{l}{}                                                                     &                                                                                                         &                                         &  &  \\ 
			Type-II           &                                             & -690.66337                                                                                                   & 5.6984                                                                                        & 5.98                                &           &                                             & \multicolumn{1}{l}{}                                                                     &                                                                                      &                             &           &                                             &                                                                                            &                                                                                                          &                                           &  &  \\ 
			Type-III          &                                             & -690.72892                                                                                          & 5.6263                                                                                    & 2.00                              &          &                                             & \multicolumn{1}{l}{}                                                                     &                                                                                    &                           &           &                                             &                                                                                           &                                                                                                          &                                           &  &  \\ 
		\end{tabular}%
	\end{ruledtabular}
\end{table*}
\subsection{Mechanical properties}
In this section, we discuss the mechanical properties of the compounds under study. Using ElaStic code \cite{elasticTool}, second-order elastic constants, $C_{ij}$, and other elastic properties are calculated. Computed values of the elastic constants, elastic modulii, and derived quantities (discussed below) are given in table \ref{tab:table_elastic1} . For cubic system, we have three independent elastic constants--$C_{11}$, $C_{12}$, and $C_{14}$. The stability of the cubic crystal can be inferred from the Born-Huang stability criteria \cite{born_huang_1956}, which is given as\\
\begin{equation}
C_{11}-2C_{12}>0\,\,\,,C_{11}>C_{12}\,\,and\,\,C_{44}>0
\end{equation}

Crystals satisfying the above conditions are considered stable. All the compounds listed in Table \ref{tab:table_elastic1} satisfy  the above-mentioned Born-Huang stability criteria and hence are mechanically stable.
Using the elastic constants, we can calculate several elastic parameters such as elastic moduli (Young's($ E $), Shear($ G $), and Bulk($ B $)), anisotropy factor($A_e $). These moduli can be used to describe the polycrystalline materials in which crystal grains are randomly oriented. We can evaluate these moduli by averaging over second-order elastic compliance ($S_{ij} $) or elastic stiffness ($C_{ij} $). The most popular averaging method is the Voigt-Ruess-Hill\cite{voigt1910lehrbuch,ruess1929,Hill_1952,HILL1963} method which can be used to calculate the elastic moduli for polycrystalline materials. 
Voigt's\cite{voigt1910lehrbuch} method assumes a uniform strain and utilize $C_{ij} $ to calculate elastic moduli but Ruess's\cite{ruess1929} method considers uniform stress and exploits $ S_{ij} $ to calculate the elastic moduli.
\par For cubic systems, the Bulk moduli is computed using Voigt and Ruess approaches, $ B_V $ and $ B_R $ respectively, are basically equal and given as
\begin{equation}
B = B_V =B_R = \frac{C_{11}+2C_{12}}{3} = \frac{1}{3(S_{11}+2S_{12})}
\end{equation}
The Shear and Bulk modulus by Voigt and Ruess, $ G_V $ and $ G_R $, are given as
\begin{equation}
G_V = \frac{C_{11}-C_{12}+3C_{44}}{5}
\end{equation}
and

\begin{equation}	
G_R = \frac{5(C_{11}-C_{12})C_{44}}{3(C_{11}-C_{12})+4C_{44}}= \frac{5}{4(S_{11}-S_{12})+3S_{44}}
\end{equation}

There is also another approach to the averaging method known as Hill's \cite{Hill_1952,HILL1963} method. In Hill's approach, Voigt and Ruess's elastic moduli are taken as the upper and lower bound. The Bulk and Shear moduli are given as
\begin{equation}
B_H= B = \frac{B_V+B_R}{2}
\end{equation}
and 
\begin{equation}
G_H = G = \frac{G_V + G_R}{2}
\end{equation}
Using this Hill's bulk (B) and shear (G) modulus, the quantities such as Pugh ratio ($k$), Poisson's ratio ($\nu$), and Young's modulus (E) are evaluated using expressions given as
\begin{equation}
k = \frac{B}{G}
\end{equation}
\begin{equation}
\nu = \frac{3B-2G}{2(3B+G)}=\frac{3k-2}{6k+2}
\end{equation}
and
\begin{equation}
E = 2G(1+\nu)
\end{equation}
\begin{figure}[ht!]
	\includegraphics[width=1\linewidth]{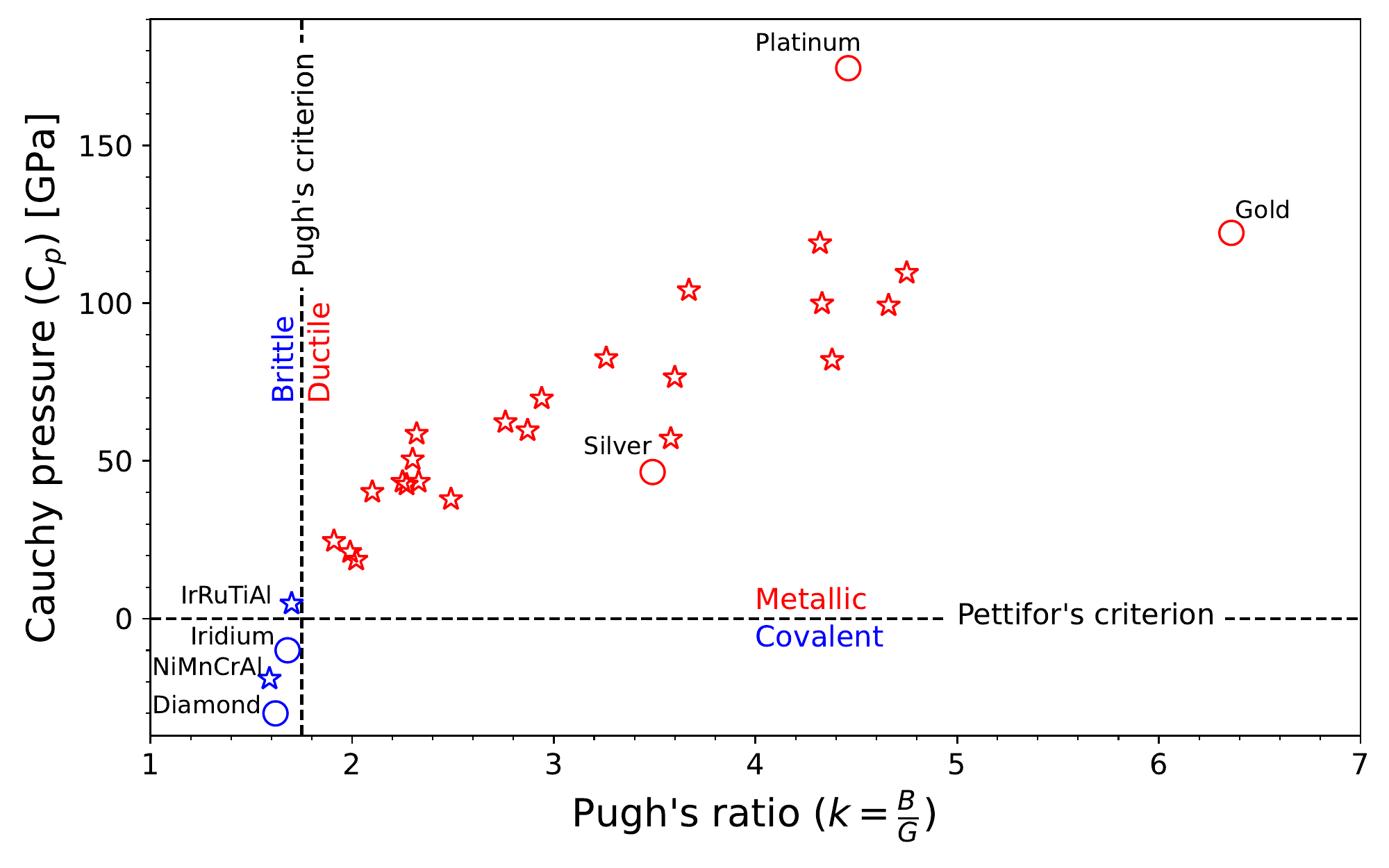}
	\caption[]{Plot of Cauchy pressure ($ C_p $) along y-axis and Pugh's ratio ($k= \frac{B}{G} $) along x-axis. Vertical and horizontal dashed lines corresponds to the Pettifor's and Pugh's criterion. Data of Diamond (which is known to be the brittle) and Gold (which is known to be most malleable) along with Iridium, Silver and Platinum is presented here for better comparison. IrCoTiAl is not shown in the figure.}
	\label{fig:ductility-brittleness}
\end{figure}

\begin{figure}[b]
	\centering
	\includegraphics[width=0.8\linewidth]{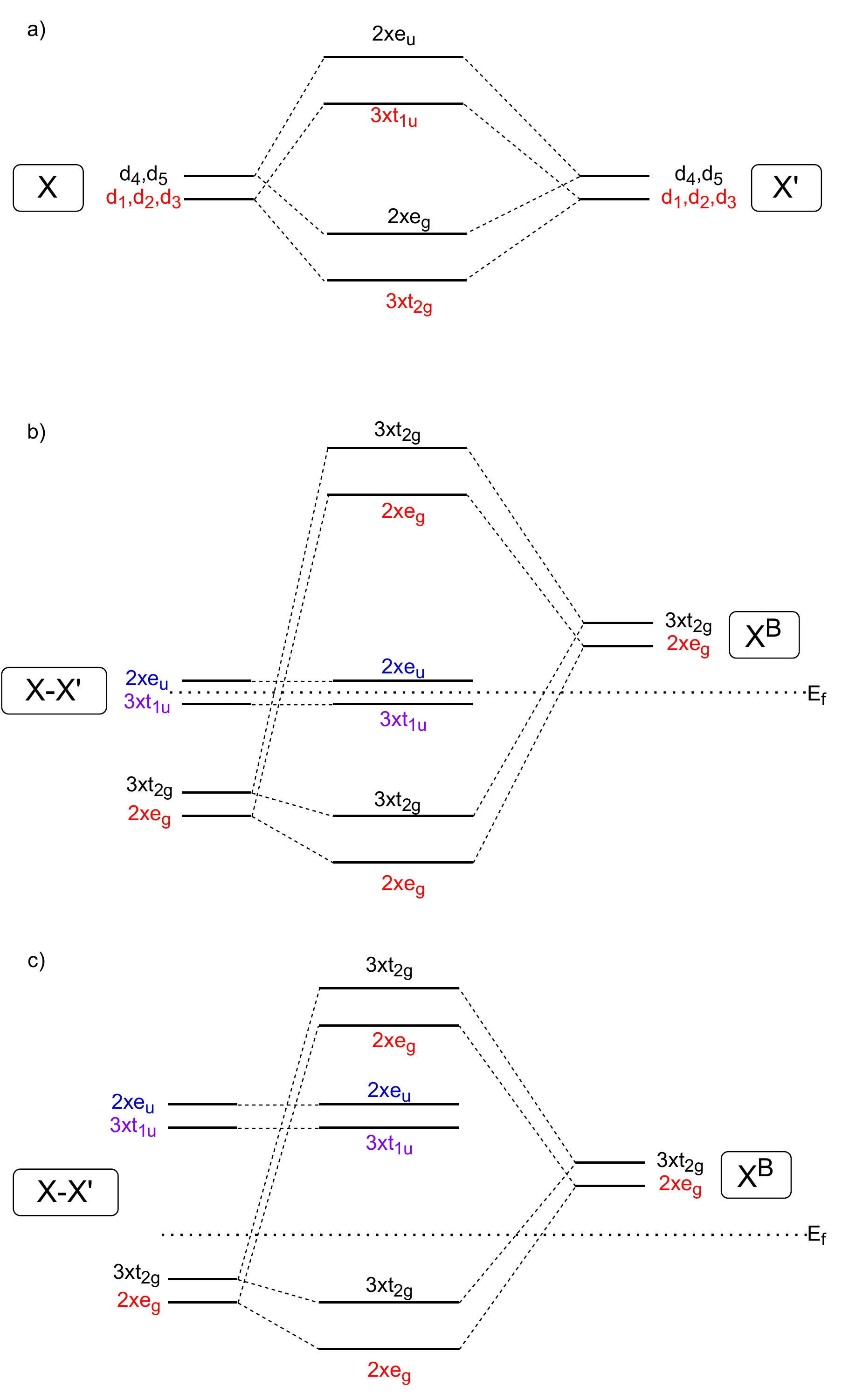}
	\caption[]{Possible hybridization schemes for spin-down bands in type-3 structures. Fig(a) represents the hybridization between X and X' atoms in A and C coordinates. Fig (b) and (c) represent the hybriziation for compounds that follow $M_t = N_v-24$ and $M_t = N_v-18$ respectively. For detailed discussion see reference \cite{ozdougan2013slater}}
	\label{fig:dos1}
\end{figure} 
The characteristic properties of compounds such as malleability, ductility, and brittleness can be studied using the Pugh ratio ($k$) and Poison's ratio ($ \nu $), whose values are given in Table \ref{tab:table_elastic1} for the compounds under study. Pugh's criterion suggests a critical value of Pugh's ratio, $ k_c =1.75$\cite{pughratio,pughcriticalelastic_Iyigor_BGValue,pughcriticalelastic_claudia,pughcriticalelastic_Seh}, which can be used to infer the ductility and brittleness of materials. A material is characterized as brittle if $ k < k_c $ and ductile if $ k > k_c $. The value of $ k$ for all the materials is greater than 1.75 except for IrCoTiAl, IrRuTiAl, and NiMnCrAl, suggesting all the materials are ductile except for these three compounds. This can be substantiated by evaluating Cauchy's Pressure ($ C_p = C_{12}-C_{44} $). A material being ductile suggests the presence of metallic bonds while brittleness of material suggests the presence of ionic or covalent bonds. The Pettifor's criterion \cite{pettifor} dictates that a positive Cauchy pressure indicates metallic bonding in material while a negative value indicates covalent or ionic bonding. Hence, the material with negative $ C_p $ can be considered brittle and that with positive $ C_p $ be considered ductile. In Table-\ref{tab:table_elastic1}, we can see all the materials except for IrCoTiAl and NiMnCrAl have positive $ C_p $, suggesting that all the materials except these two are ductile. This observation complies with our previous prediction of ductility and brittleness based on the value of Pugh's ratio ($ k $).From Figure \ref{fig:ductility-brittleness} and Table-\ref{tab:table_elastic1},  one can notice that compound IrCoTiAl and NiMnCrAl, both lie below Pettifor's criterion and to the left of Pugh's criterion. Hence, we can say that IrCoTiAl and NiMnCrAl both are likely to be brittle but the behavior of IrRuTiAl lies on the borderline of brittleness and ductility, when one compares with the data of diamond, iridium (which are brittle) and gold, silver, and platinum (which are ductile).
\par  The anisotropy factor ($ A_e $) is another parameter for describing mechanical stability. For the cubic system, anisotropy factor is given as
\begin{equation}
A_e = \frac{2C_{44}}{C_{11}-C_{12}}
\end{equation}
It is a known fact that for isotropic materials, $ A_e $ is equal to 1, and materials that possess very high anisotropy have the tendency to deviate from the cubic structure occasionally. The materials that have $ A_e <0$ violate at least one of the Born-Huang stability criteria and hence are expected to be mechanically unstable. Many of the compounds under study, as can be seen in Table\ref{tab:table_elastic1}, have $ A_e>1 $ and therefore suggest some anisotropy. The compound IrFeZrAl has anisotropy factor equal to unity implying that it is an isotropic material. Recently Felser and their team studied different elastic properties of many Heusler compounds\cite{pughcriticalelastic_claudia}. Among them is Co$_2$CrAl, an experimentally synthesized full Huesler compound that is known to crystallize in stable cubic structure\cite{hakimi2010structural,dubowik2007structure}. The A$_e$ value for this compound is 3.28. From Felser's results one can safely conclude that values close to or below 3.28 means that cubic structure is stable. Even larger values mean that the deviation from cubic structures is very small, the c/a ratio is close to 1, and the electronic and magnetic properties are almost identical to the cubic phase. We note that only three compounds, CoRuCrGe, CoRuCrSi, and NiFeVAl have large value of A$_e$ and one can expect some deformation from cubic lattice structure in these compounds.     
\begin{table*}[t!]
	\caption{Calculated values of elastic constants ($ C_{ij} $ in GPa), Bulk modulus ($ B $ in GPa), Shear modulus ($ G $ in GPa), Young's modulus (E in GPa), Pugh's ratio ($ k = B/G$), Poisson's ration ($ \nu $), Cauchy pressure ($ C_p $) and Anisotropic factor ($ A_e $) for the lowest energy structure.}
	\label{tab:table_elastic1}
\begin{ruledtabular}
		\begin{tabular}{lllllllllll}
			Compounds & C$_{11}$ & C$_{12}$ & C$_{44}$ & $B$    & $G$    & $ E $      & $k$ & $\nu$ & $C_p$ & $A_e$ \\ \bottomrule[0.05em]
			CoCrZrAl  & 158.6    & 128.7    & 71.6     & 138.66 & 38.70   & 106.21 & 3.58    & 0.37 & 57.09               & 4.78  \\ 
			CoIrMnSb  & 195.2    & 150.4    & 73.9     & 165.33 & 45.89  & 126.03 & 3.60    & 0.37 & 76.50               & 3.29  \\ 
			CoIrMnSn  & 221.4    & 159.4    & 99.7     & 180.06 & 62.53  & 168.14 & 2.87    & 0.34 & 59.70               & 3.21  \\ 
			CoRuCrGa  & 259.3    & 185.9    & 123.6    & 210.36 & 76.15  & 203.87 & 2.76    & 0.33 & 62.30               & 3.36  \\ 
			CoRuCrGe  & 230.7    & 210.4    & 111.1    & 217.16 & 46.51  & 130.25 & 4.66    & 0.40 & 99.30               & 10.94 \\
			CoRuCrSi  & 250.4    & 229.2    & 119.6    & 236.26 & 49.69  & 139.31 & 4.75    & 0.40 & 109.60              & 11.28 \\ 
			CoRuZrSi  & 242.4    & 175.1    & 56.1     & 197.53 & 45.70   & 127.28 & 4.32    & 0.39 & 119.00              & 1.66  \\
			FeRuCrGe  & 321.4    & 167.3    & 127.1    & 218.66 & 103.98 & 269.26 & 2.10    & 0.29 & 40.20               & 1.64  \\ 
			IrCoTiAl  & 384.8    & 27.6     & 141.6    & 146.66 & 155.39 & 344.51 & 0.94    & 0.10 & -114.00             & 0.79  \\ 
			IrFeZrAl  & 327.2    & 125.9    & 101.3    & 193.00    & 101.03 & 258.08 & 1.91    & 0.27 & 24.60               & 1.00  \\
			IrMnCrGe  & 256.4    & 145.8    & 103.3    & 182.66 & 80.38  & 210.31 & 2.27    & 0.30 & 42.50               & 1.86  \\
			IrMnCrSi  & 291.5    & 167.0    & 116.5    & 208.50  & 90.59  & 237.39 & 2.30    & 0.31 & 50.50               & 1.87  \\ 
			IrRuTiAl  & 410.5    & 135.2    & 130.4    & 226.96 & 133.25 & 334.33 & 1.70    & 0.25 & 4.79                & 0.94  \\ 
			NiFeVAl   & 214.2    & 196.3    & 114.3    & 202.26 & 46.09  & 128.51 & 4.38    & 0.39 & 82.00               & 12.77 \\ 
			NiMnCrAl  & 212.6    & 106.9    & 125.9    & 142.13 & 88.87  & 220.64 & 1.59    & 0.24 & -19.00              & 2.38  \\ 
			RhCoZrAl  & 258.3    & 120.3    & 76.9     & 166.30  & 73.63  & 192.49 & 2.25    & 0.30 & 43.39               & 1.11  \\ 
			RhCrTiAl  & 209.0    & 143.1    & 105.2    & 165.06 & 66.17  & 175.11 & 2.49    & 0.32 & 37.89               & 3.19  \\ 
			RhFeMnGe  & 248.6    & 189.4    & 106.8    & 209.13 & 64.09  & 174.46 & 3.26    & 0.36 & 82.60               & 3.60  \\ 
			RhFeMnSi  & 317.6    & 174.4    & 115.8    & 222.13 & 95.49  & 250.57 & 2.32    & 0.31 & 58.60               & 1.61  \\ 
			RhFeTiGe  & 216.4    & 181.4    & 81.5     & 193.06 & 44.49  & 123.96 & 4.33    & 0.39 & 99.90               & 4.65  \\ 
			RhFeTiSi  & 251.5    & 187.4    & 83.3     & 208.76 & 56.80   & 156.23 & 3.67    & 0.37 & 104.10              & 2.59  \\ 
			RuCrTiSi  & 249.5    & 181.8    & 112.0    & 204.36 & 69.48  & 187.23 & 2.94    & 0.34 & 69.80               & 3.30  \\ 
			RuCrZrGa  & 231.8    & 118.1    & 97.0     & 156.00   & 78.28  & 201.20  & 1.99    & 0.28 & 21.09               & 1.70  \\ 
			RuMnCrSi  & 249.7    & 151.1    & 107.7    & 183.96 & 78.70   & 206.65 & 2.33    & 0.31 & 43.39               & 2.18  \\ 
			CoMnCrSi  & 256.0    & 148.7    & 130.0    & 184.46 & 91.15   & 234.78 & 2.02    & 0.28 & 18.69               & 2.42  \\ 
		\end{tabular}%
	\end{ruledtabular}
\end{table*}
\begin{figure}[ht!]
	\centering
	\includegraphics[width=1\linewidth]{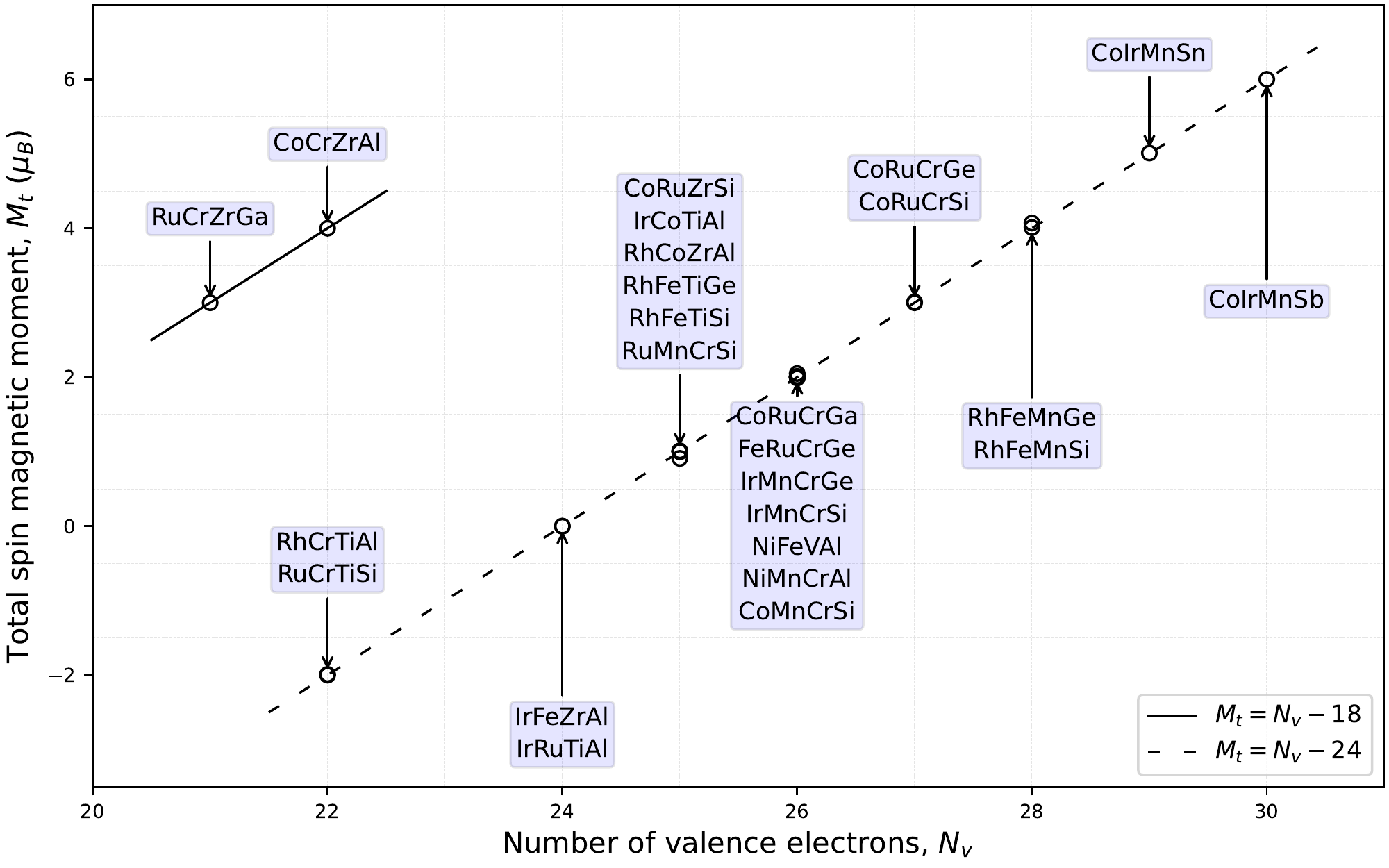}
	\caption[]{Calculated total spin moments (in $\mu_B$) for all the studied Heusler compounds as a function of total number of valence electrons. The solid and dashed line represent different forms of Slater-Pauling rule; $M_t = N_v-18$ and $M_t = N_v-24$ respectively. }
	\label{fig:sp}
\end{figure} 

\subsection{Slater-Pauling behavior and hybridization in quaternary Heusler alloys}
One of the most interesting properties of Heusler compounds is probably the Slater-Pauling rule  which allows one to predict the total spin magnetic moment, $M_t$, of the compound by knowing the total number of valence electrons, $N_v$ (see Figure \ref{fig:sp}). The perfect half-metallic ferromagnets and spin gapless semiconductors follow this rule strictly in Heusler compounds due to their integer value of the total spin magnetic moment. The valence electrons in any Heusler compounds can be either in spin-up or spin-down states due to which the difference in their number gives the observed value of the total spin magnetic moment. In Quaternary Heusler compounds, the relative position of \textit{d}-states of Y atoms with respect to X and X\textquotesingle$\,$can often lead to a complex Slater-Pauling rule.  Among 25 Heusler compounds under study, all of the compounds obey  $M_t = N_v-24$ rule except two, CoCrZrAl and RuCrZrGa. In these compounds, $t_{1u}$ states are relatively higher in energy and hence are not occupied, leading to modified $M_t = N_v-18$ relations.

\par   The hybridization scheme for different types of Heusler alloys are well known\cite{Galanakis2002,Galanakisinverse,ozdougan2013slater} and has been used extensively to explain the observed electronic and magnetic properties of a large number of compounds \cite{dhakal2021prediction,our, our2}. In Quaternary Heusler Compounds which follow the $M_t = N_v-24$ rule, due to the equivalent nature of A and C coordinates, X and X\textquotesingle$\,$elements hybridize first to form double $e_g$ and triple $t_{2g}$ hybrids. The Y transition metal at D coordinates in turn hybridizes with these hybrid states to form 5 bonding, 5 anti-bonding and 5 non-bonding \textit{d}-states. The main group element(Z) contributes one \textit{s}, and triple degenerated \textit{p} bands which are relatively lower in energy compared to \textit{d} bands and contribute to the stability of the structure by decreasing the effective \textit{d}-charge concentration.  The relative position of non-bonding \textit{d}-hybrids, which consist of three occupied $t_{1u}$ and two unoccupied $e_u$ states, is of our interest as it determines the energy gap of the spin-down band in Quaternary Heusler alloys. The same scheme is true for Quaternary Heusler compounds that follow $M_t = N_v-18$ rule, except that in these compounds $t_{1u}$ states are unoccupied. For the compounds which crystallize in type-1 structure, X at B and Y at D site forms the octahedral symmetry and hence hybridize first to form different states which in turn hybridize with X\textquotesingle$\,$ at C site to form hybridized states as explained above. Of course in the case of the type-1 and type-3 structures, since the hybridization scheme arises from different atoms, the position of the resulting bands would be different. It is also important to note that because of the symmetry of the 216 space group one can exchange the position of X and X\textquotesingle$\,$, and Y and Z atoms without altering the structure of the crystal.

\subsection{Electronic and magnetic properties}                  
In Table \ref{tab:my-table}, we gather calculated spin magnetic moments of the compounds along with optimized lattice parameters. Almost all compounds show an integer value for the  total magnetic moment in the stable structure, a prerequisite for compounds to be half-metallic. In our convention, the gap remains in the spin-down band which occupies twelve electrons in all the cases so that if the compounds have less than 24 valence electrons and follow $M_t = N_v-24$ rule, the total magnetic moment would be negative. In this section, we discuss the electronic and magnetic properties of the compounds by dividing them into three categories: i) compounds that follow $M_t = N_v-24$ rule ii) compounds that obey $M_t = N_v-18$ relation and iii) compounds that crystallize in type-1 structure.
\begin{figure}[b]
	\centering
	\includegraphics[width=1\linewidth]{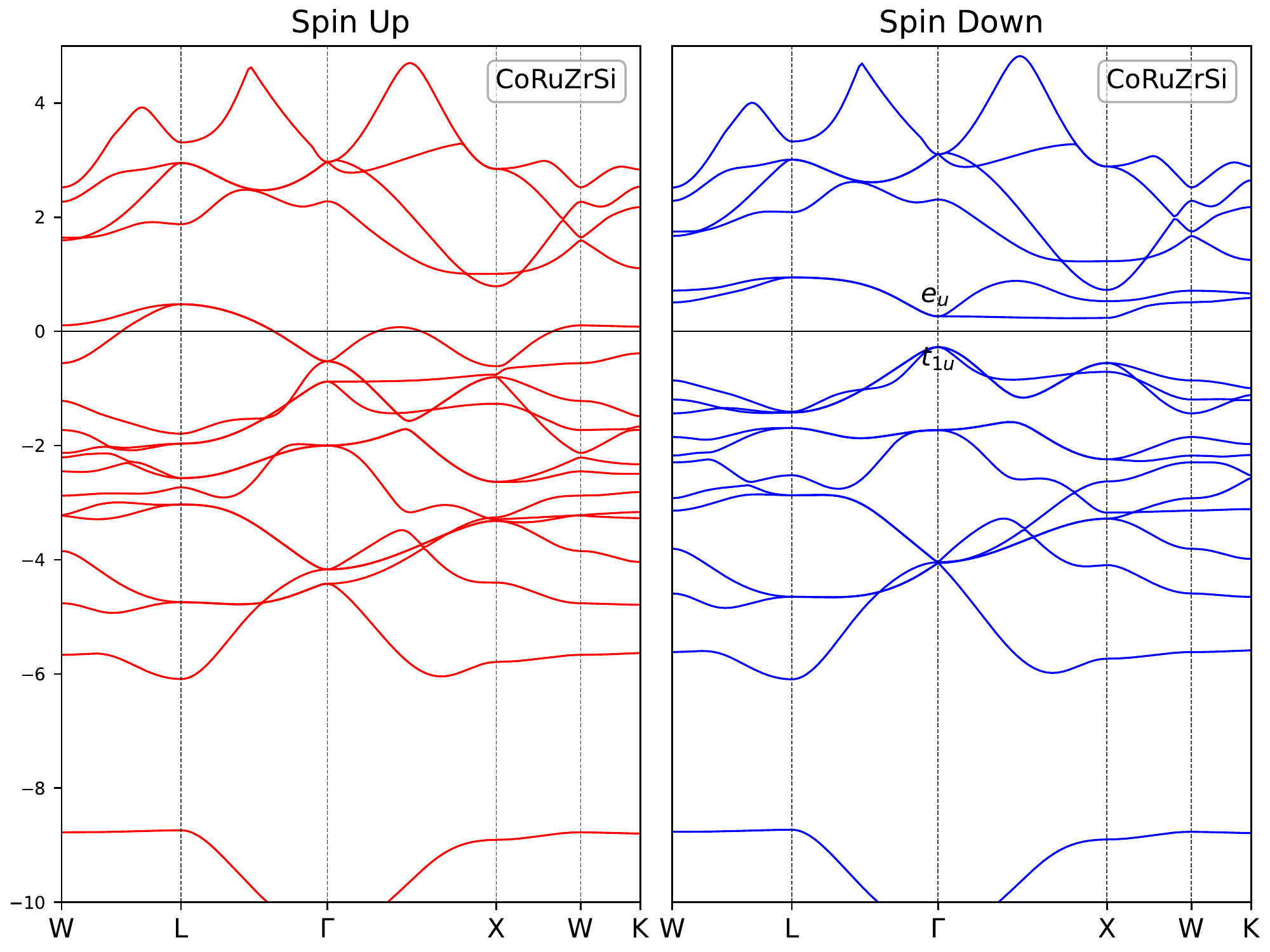}
	\caption[]{Calculated spin-resolved band structure of CoRuZrSi using optimized lattice parameter of 6.1191\AA. The red and blue color represent the spin-up and spin-down bandstructures respectively.}
	\label{fig:band1}
\end{figure} 
\begin{figure}[ht!]
	\centering
	\includegraphics[width=1\linewidth]{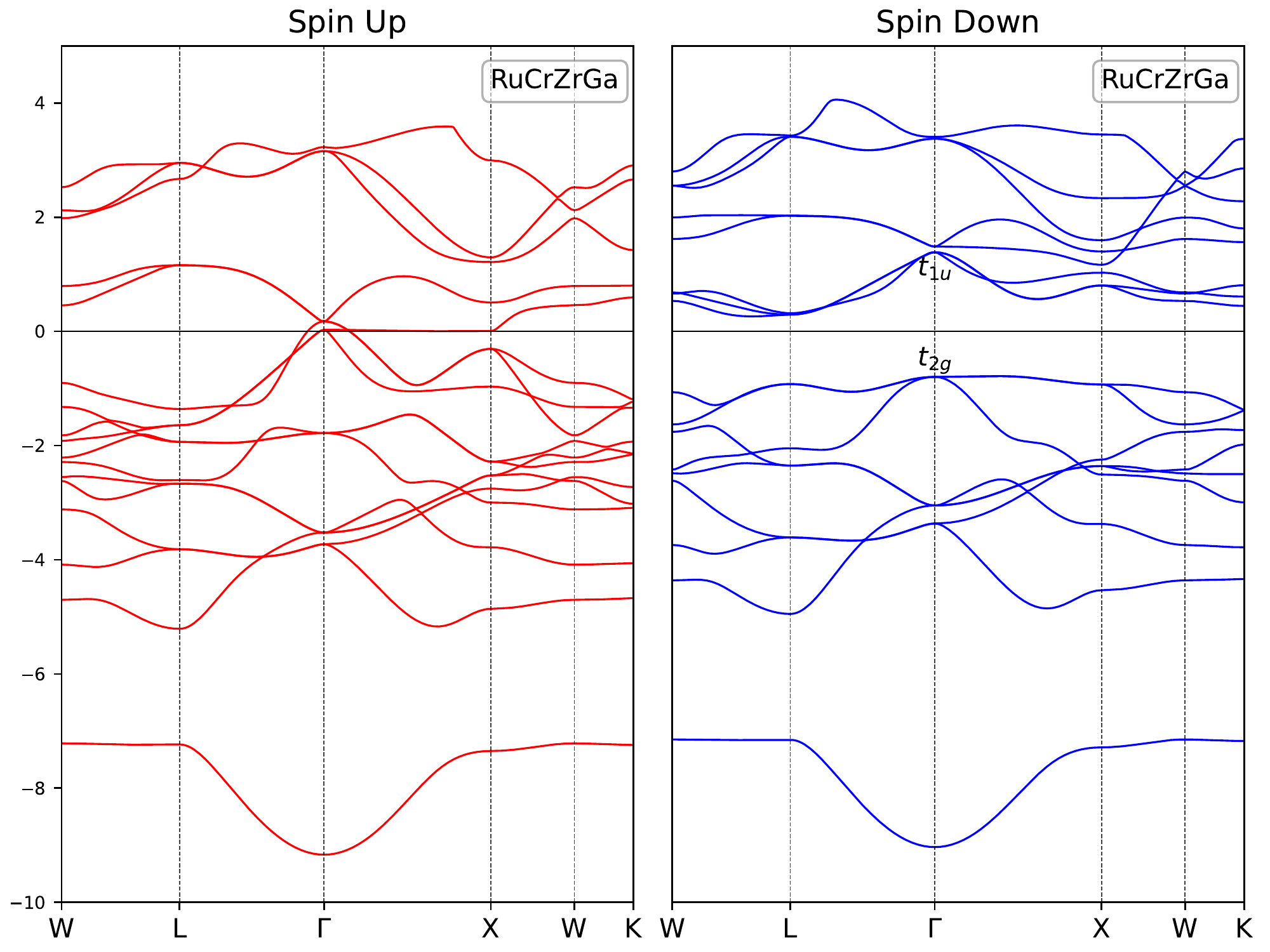}
	\caption[]{Calculated spin-resolved band structure of RuCrZrGa using optimized lattice parameter of 6.3195\AA. The red and blue color represent the spin-up and spin-down bandstructures respectively.}
	\label{fig:band2}
\end{figure} 
\begin{figure}[ht!]
	\centering
	\includegraphics[width=1\linewidth]{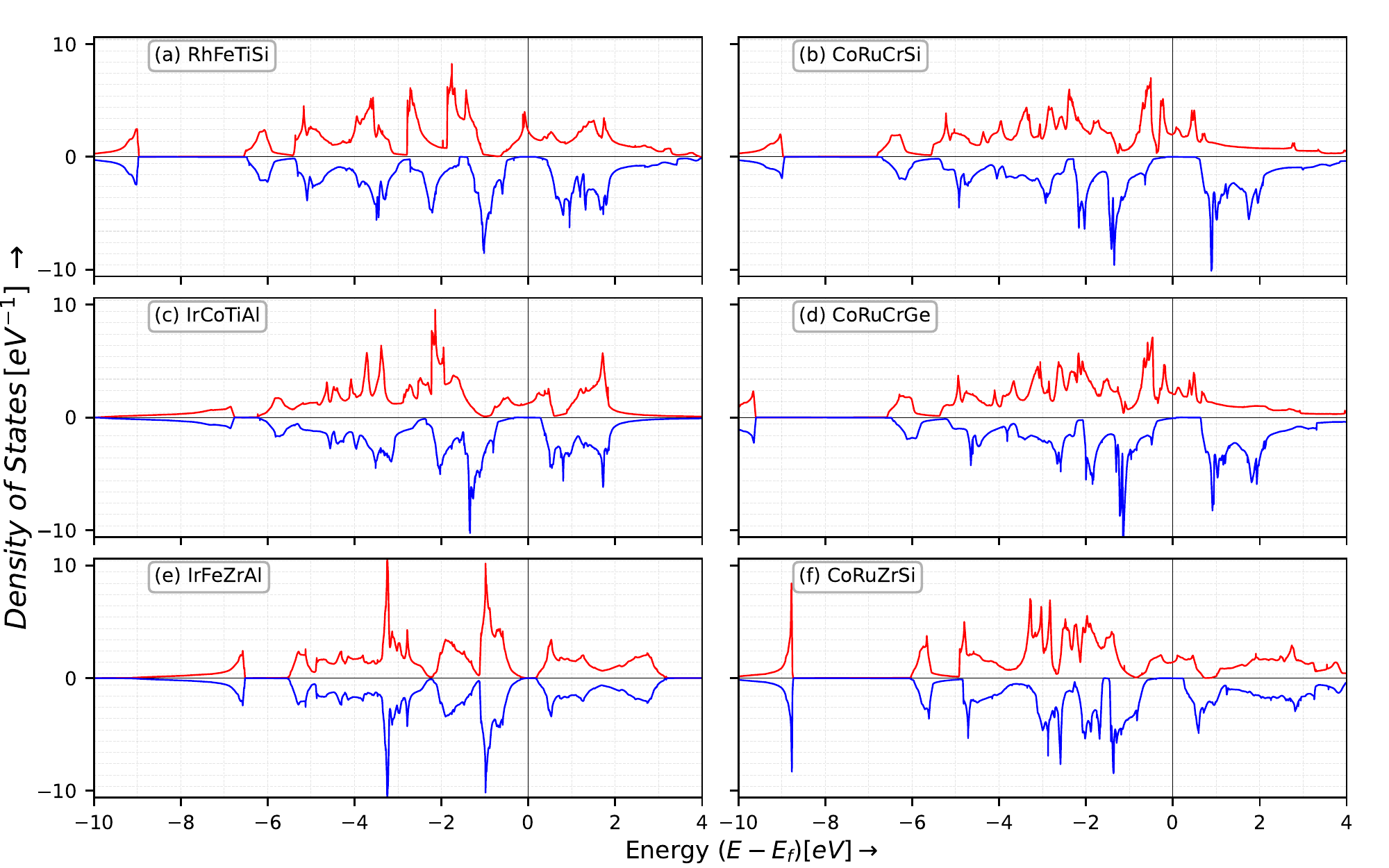}
	\caption[]{Spin-polarized total density of states of the selected compounds. The red and blue color represent spin up and spin-down states respectively. The Fermi level is at the zero of the energy axis.}
	\label{fig:dos1}
\end{figure}

\begin{figure}[ht!]
	\centering
	\includegraphics[width=1\linewidth]{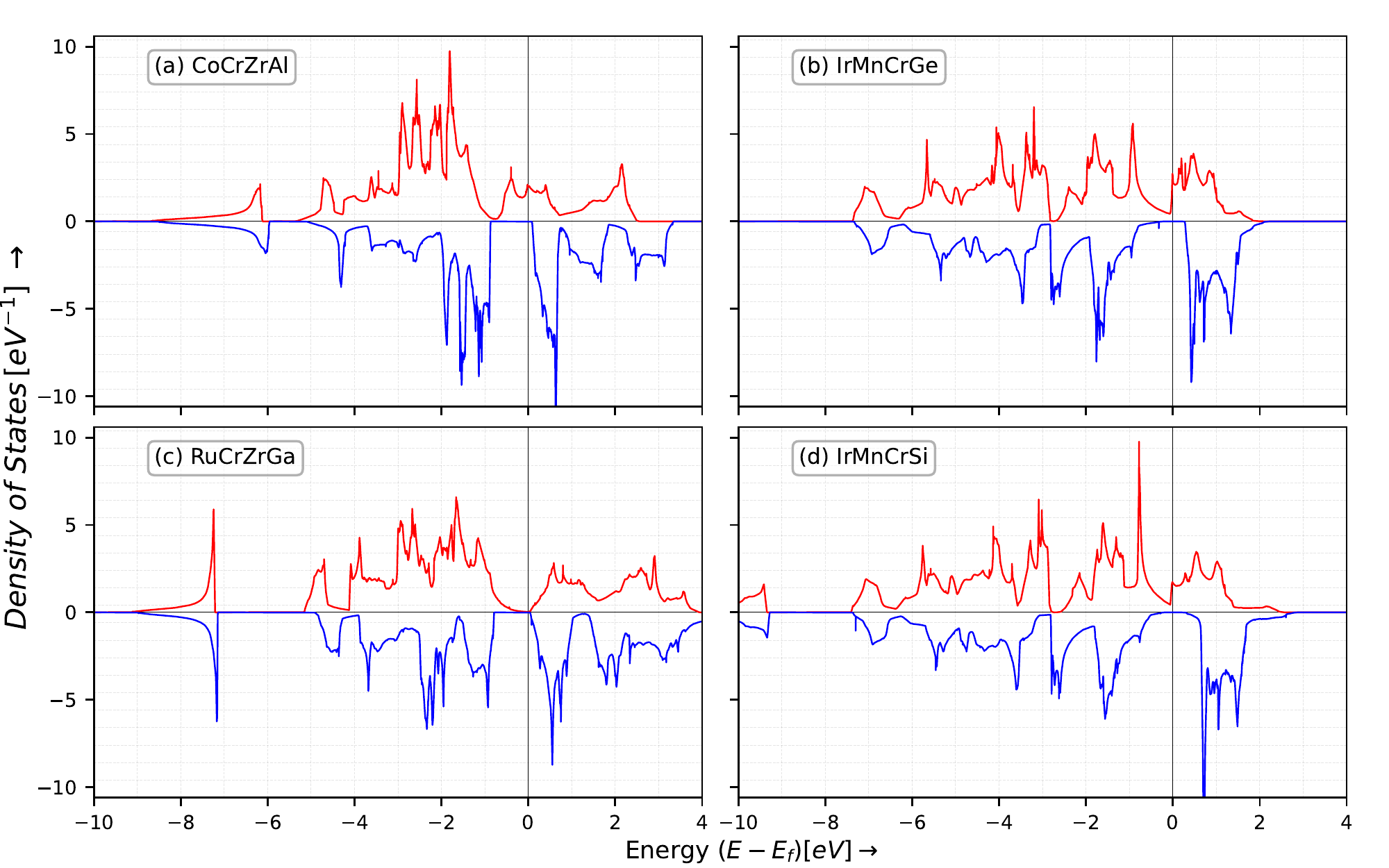}
	\caption[]{Spin-polarized total density of states of the selected compounds. The red and blue color represent spin up and spin-down states respectively. The Fermi level is at the zero of the energy axis.}
	\label{fig:dos2}
\end{figure} 
\par In Figure \ref{fig:band1}, we present spin resolved bandstructure of CoRuZrSi,as a representative of the compounds that follow the $M_t = N_v-24$ rule.   One can observe the usual metallic behavior on the spin-up channel whereas the Fermi-level lies in the gap for the spin-down channel resulting in a desired half-metallic gap. The lower lying \textit{s} and \textit{p} bands, which are similar in both spin channels, do not have a significant contribution on the half-metallic gap. The gap arises from the non-overlapping nature of $t_{1u}$ and $e_u$ states which are localized around Co and Ru. These states can not couple with Zr since there are no \textit{d} states on Zr that transform through \textit{u} representation. As Co and Ru are the next neighboring atoms the energy distance between $t_{1u}$ and $e_u$ is small resulting in a smaller gap, a typical property that is observed in Quaternary Heulser alloys. The bandstructures for all of the compounds is studied and can be found in the supplementary section. In Figure \ref{fig:dos1}, we gather the total Density of States (DOS) of the selected compounds under this category. Let us first compare CoRuCrSi and CoRuCrGe (Fig.\ref{fig:dos1}(b,d)). The former has a larger half-metallic gap while the latter compound is nearly half-metallic since on can observe very small spin-down DOS at the Fermi level. This is expected since Si has its p-states higher in energy than Ge and thus closer to the gap. This is even more intense when we compare Al with Ga. The small admixture of the p-states in the bands just below the Fermi level leads to an opening of the gap since the p-d hybridization becomes more important. The same occurs also in the usual semiconductors. When we compare CoRuCrSi with CoRuZrSi (Fig.\ref{fig:dos1}(b,f)) one can observe a large half-metallic gap on both compounds but the bandwidth of the spin-up bandstructure in CoRuZrSi is larger than CoRuZrSi. This is because of the fact that Zr has two valence electrons less than Cr, and also Zr is a 4d metal. This is also the reason why there is an important part of it unoccupied in CoRuZrSi. Among the other three compounds, RhFeTiSi and IrCoTiAl (Fig.\ref{fig:dos1}(a,c)) are half-metallic in nature with a large gap while IrFeZrAl (Fig.\ref{fig:dos1}(e)) is a special case. The gap is in both spin directions, and the gap lies in the same energy region. Also, there is a striking similarity in DOS in both spin directions. Thus it can be categorized as non-magnetic semiconductor which can be confirmed from the observed total magnetic moment of the compound in Table \ref{tab:table_elastic}.          
\begin{table}[b]
	\caption{Comparison of spin mangetic moments of type-1 and type-3 structure of the compounds whose stable structure is type-1.}
	\label{tab:table5}
	\begin{ruledtabular}
		\begin{tabular}{lccccccc}
			\multicolumn{1}{l}{\multirow{2}{*}{Compounds}} & \multicolumn{1}{l}{\multirow{2}{*}{N$_V$}} & \multicolumn{1}{c}{\multirow{2}{*}{\begin{tabular}[c]{@{}c@{}}a$_{opt}$\\ (\AA)\end{tabular}}} & \multicolumn{5}{c}{Magnetic Moments ($\mu_B$)}                                                                                 \\ \cline{4-8} 
			\multicolumn{1}{l}{}                           & \multicolumn{1}{l}{}                       & \multicolumn{1}{c}{}                                                                                           & \multicolumn{1}{c}{m$_{tot}$} & \multicolumn{1}{c}{m$_{X}$} & \multicolumn{1}{c}{m$_{X^\prime}$} & \multicolumn{1}{c}{m$_{Y}$} & m$_{Z}$ \\ \bottomrule[0.05em]
			\multicolumn{8}{c}{Type-1 (Stable)}                                                                                                                                                                                                                                                                                                                             \\ \bottomrule[0.05em]
			\multicolumn{1}{l}{IrMnCrGe}                   & \multicolumn{1}{c}{26}                     & \multicolumn{1}{c}{6.0098}                                                                                     & \multicolumn{1}{c}{2.00}               & \multicolumn{1}{c}{0.24}  & \multicolumn{1}{c}{3.22}  & \multicolumn{1}{c}{-1.54} & 0.04  \\ 
			\multicolumn{1}{l}{IrMnCrSi}                   & \multicolumn{1}{c}{26}                     & \multicolumn{1}{c}{5.8909}                                                                                     & \multicolumn{1}{c}{2.00}               & \multicolumn{1}{c}{0.20}  & \multicolumn{1}{c}{3.00}  & \multicolumn{1}{c}{-1.26} & 0.03  \\ 
			\multicolumn{1}{l}{NiMnCrAl}                   & \multicolumn{1}{c}{26}                     & \multicolumn{1}{c}{5.8250}                                                                                     & \multicolumn{1}{c}{2.00}               & \multicolumn{1}{c}{0.64}  & \multicolumn{1}{c}{2.90}  & \multicolumn{1}{c}{-1.49} & -0.02 \\ 
			\multicolumn{1}{l}{RuMnCrSi}                   & \multicolumn{1}{c}{25}                     & \multicolumn{1}{c}{5.8742}                                                                                     & \multicolumn{1}{c}{1.01}               & \multicolumn{1}{c}{-0.03} & \multicolumn{1}{c}{2.88}  & \multicolumn{1}{c}{-1.80} & 0.03  \\ 
			\multicolumn{1}{l}{CoMnCrSi}                   & \multicolumn{1}{c}{26}                     & \multicolumn{1}{c}{5.6796}                                                                                     & \multicolumn{1}{c}{2.01}               & \multicolumn{1}{c}{0.91}  & \multicolumn{1}{c}{2.57}  & \multicolumn{1}{c}{-1.35} & -0.02 \\ \bottomrule[0.05em]
			\multicolumn{8}{c}{Type-3 (Unstable)}                                                                                                                                                                                                                                                                                                                                      \\ \bottomrule[0.05em]
			\multicolumn{1}{l}{IrMnCrGe}                   & \multicolumn{1}{c}{26}                     & \multicolumn{1}{c}{5.9509}                                                                                     & \multicolumn{1}{c}{2.01}               & \multicolumn{1}{c}{0.09}  & \multicolumn{1}{c}{-0.65} & \multicolumn{1}{c}{2.35}  & 0.03  \\ 
			\multicolumn{1}{l}{IrMnCrSi}                   & \multicolumn{1}{c}{26}                     & \multicolumn{1}{c}{5.8485}                                                                                     & \multicolumn{1}{c}{2.00}               & \multicolumn{1}{c}{0.08}  & \multicolumn{1}{c}{-0.46} & \multicolumn{1}{c}{2.19}  & 0.02  \\ 
			\multicolumn{1}{l}{NiMnCrAl}                   & \multicolumn{1}{c}{26}                     & \multicolumn{1}{c}{5.7750}                                                                                     & \multicolumn{1}{c}{1.88}               & \multicolumn{1}{c}{0.67}  & \multicolumn{1}{c}{-1.10} & \multicolumn{1}{c}{2.17}  & -0.01 \\ 
			\multicolumn{1}{l}{RuMnCrSi}                   & \multicolumn{1}{c}{25}                     & \multicolumn{1}{c}{5.7837}                                                                                     & \multicolumn{1}{c}{1.01}               & \multicolumn{1}{c}{-0.03} & \multicolumn{1}{c}{-0.53} & \multicolumn{1}{c}{1.46} & 0.02  \\
			\multicolumn{1}{l}{CoMnCrSi}                   & \multicolumn{1}{c}{26}                     & \multicolumn{1}{c}{5.6263}                                                                                     & \multicolumn{1}{c}{2.00}               & \multicolumn{1}{c}{1.05}  & \multicolumn{1}{c}{-0.72} & \multicolumn{1}{c}{1.66}  &   -0.03      \\
		\end{tabular}
		\end{ruledtabular}
\end{table}
 
\par We have two compounds, CoCrZrAl and RuCrZrGa, which obey  $M_t = N_v-18$ rule.  We present the spin-resolved bandstructure of RuCrZrGa in the Figure \ref{fig:band2}.  The hybridization scheme and the band structure are similar as discussed above except that now the triple degenerated  $t_{1u}$ states are above the Fermi-level so that the gap lies in between   $t_{1u}$ and  $t_{2g}$ states. The state counting suggests that there are 9 minority states in the spin-down channel and 12 majority states in the spin-up channel with a difference of 3, which appears as the total magnetic moment (3 $\mu_B$) of RuCrZrGa. In Figure \ref{fig:dos2}(a,c), we present the DOS of CoCrZrAl and RuCrZrGa. The DOS of CoCrZrAl confirms the half-metallic nature of the compound while the valley approaching peak can be observed on the spin-up states of RuCrZrGa. There is a usual gap on the spin-down state. We categorize this compound as a nearly spin-gapless semiconductor(SGS) since from spin-resolved bandstructure (see 
Figure \ref{fig:band2}) we observe a small crossing of bands at $\Gamma$ point preventing it from being perfect SGS compound.    

\par We have five compounds XMnCrZ(X = Ir,Ni,Ru,Co and Z = Ge,Si,Al) that prefer to crystallize in type-1 structure. All of these compounds follow the $M_t = N_v-24$ rule for the observed total spin magnetic moment. To find out the cause of the observed anomaly, we analyze the projected density of states(PDOS) of both type-1 and type-3 structure(not presented here) and the spin-resolved individual magnetic moment of the constituent elements (see Table \ref{tab:table5}). The total magnetic moment is identical in both cases except in NiMnCrAl where there is a small difference of 0.12$\mu_B$. In both cases, the individual spin magnetic moment of X and Z atoms carry small values of spin magnetic moments. In both type-1 and type-3 compounds the Mn and Cr are the nearest neighbors and thus have anti-parallel spin magnetic moments as expected  by the semi-empirical Bethe-Slater curve. The compounds are half-metal in both structure. This is expected since half-metallicity when feasible lower the total energy, and Cr and Mn atoms have just one valence electron difference. Thus their exchange of sites along the diagonal does not alter the electronic band picture significantly.
\begin{table}[b]
	\caption{Calculated spin magnetic moments for stable structure (Type-3).}
	\label{tab:table_elastic}
	\begin{ruledtabular}
		\begin{tabular}{lccccccc}
			\multirow{2}{*}{Compounds} & \multicolumn{1}{l}{\multirow{2}{*}{N$_V$}} & \multirow{2}{*}{\begin{tabular}[c]{@{}c@{}}a$_{opt}$\\ (\AA)\end{tabular}} & \multicolumn{5}{c}{Magnetic Moments ($\mu_B$)}                                                                                  \\ \cline{4-8} 
			& \multicolumn{1}{l}{}                       &                                                                                           & \multicolumn{1}{c}{m$_{tot}$} & \multicolumn{1}{c}{m$_{X}$} & \multicolumn{1}{c}{m$_{X\textquotesingle}$} & \multicolumn{1}{c}{m$_{Y}$} & m$_{Z}$ \\ \bottomrule[0.05em]
			CoCrZrAl                   & 22                                          & 6.2422                                                                                    & \multicolumn{1}{c}{4.00}               & \multicolumn{1}{c}{1.04}  & \multicolumn{1}{c}{2.97}   & \multicolumn{1}{c}{-0.10} & -0.04 \\ 
			CoIrMnSb                   & 30                                          & 6.2390                                                                                    & \multicolumn{1}{c}{6.00}               & \multicolumn{1}{c}{1.64}  & \multicolumn{1}{c}{0.52}   & \multicolumn{1}{c}{3.68}  & 0.04  \\ 
			CoIrMnSn                   & 29                                          & 6.1999                                                                                    & \multicolumn{1}{c}{5.01}               & \multicolumn{1}{c}{1.38}  & \multicolumn{1}{c}{0.27}   & \multicolumn{1}{c}{3.44}  & -0.03 \\ 
			CoRuCrGa                   & 26                                          & 5.8747                                                                                    & \multicolumn{1}{c}{2.05}               & \multicolumn{1}{c}{0.79}  & \multicolumn{1}{c}{-0.31}  & \multicolumn{1}{c}{1.63}  & -0.04 \\ 
			CoRuCrGe                   & 27                                          & 5.8992                                                                                    & \multicolumn{1}{c}{3.01}               & \multicolumn{1}{c}{1.15}  & \multicolumn{1}{c}{-0.08}  & \multicolumn{1}{c}{1.94}  & -0.03 \\ 
			CoRuCrSi                   & 27                                          & 5.7930                                                                                    & \multicolumn{1}{c}{3.00}               & \multicolumn{1}{c}{1.17}  & \multicolumn{1}{c}{-0.02}  & \multicolumn{1}{c}{1.85}  & -0.04 \\ 
			CoRuZrSi                   & 25                                          & 6.1191                                                                                    & \multicolumn{1}{c}{1.00}               & \multicolumn{1}{c}{0.93}  & \multicolumn{1}{c}{0.20}   & \multicolumn{1}{c}{-0.05} & 0.00  \\ 
			FeRuCrGe                   & 26                                          & 5.8872                                                                                    & \multicolumn{1}{c}{2.01}               & \multicolumn{1}{c}{-0.07} & \multicolumn{1}{c}{-0.13}  & \multicolumn{1}{c}{2.07}  & 0.00  \\ 
			IrCoTiAl                   & 25                                          & 6.0212                                                                                    & \multicolumn{1}{c}{1.00}               & \multicolumn{1}{c}{0.12}  & \multicolumn{1}{c}{1.07}   & \multicolumn{1}{c}{-0.07} & -0.02 \\ 
			IrFeZrAl                   & 24                                          & 6.2182                                                                                    & \multicolumn{1}{c}{0.00}               & \multicolumn{1}{c}{-0.00} & \multicolumn{1}{c}{0.00}   & \multicolumn{1}{c}{0.00}  & 0.00  \\ 
			IrRuTiAl                   & 24                                          & 6.1029                                                                                    & \multicolumn{1}{c}{0.00}               & \multicolumn{1}{c}{0.00}  & \multicolumn{1}{c}{0.00}   & \multicolumn{1}{c}{0.00}  & 0.00  \\ 
			NiFeVAl                    & 26                                          & 5.7846                                                                                    & \multicolumn{1}{c}{1.99}               & \multicolumn{1}{c}{0.68}  & \multicolumn{1}{c}{0.93}   & \multicolumn{1}{c}{0.38}  & -0.01 \\
			RhCoZrAl                   & 25                                          & 6.2225                                                                                    & \multicolumn{1}{c}{0.91}               & \multicolumn{1}{c}{0.20}  & \multicolumn{1}{c}{0.87}   & \multicolumn{1}{c}{-0.06} & -0.01 \\
			RhCrTiAl                   & 22                                          & 6.0996                                                                                    & \multicolumn{1}{c}{-2.00}              & \multicolumn{1}{c}{-0.08} & \multicolumn{1}{c}{-2.04}  & \multicolumn{1}{c}{0.25}  & 0.02  \\
			RhFeMnGe                   & 28                                          & 5.9089                                                                                    & \multicolumn{1}{c}{4.07}               & \multicolumn{1}{c}{0.26}  & \multicolumn{1}{c}{0.80}   & \multicolumn{1}{c}{3.05}  & -0.04 \\
			RhFeMnSi                   & 28                                          & 5.8021                                                                                    & \multicolumn{1}{c}{4.01}               & \multicolumn{1}{c}{0.29}  & \multicolumn{1}{c}{0.81}   & \multicolumn{1}{c}{2.95}  & -0.05 \\ 
			RhFeTiGe                   & 25                                          & 6.0132                                                                                    & \multicolumn{1}{c}{1.01}               & \multicolumn{1}{c}{0.19}  & \multicolumn{1}{c}{1.16}   & \multicolumn{1}{c}{-0.21} & -0.01 \\
			RhFeTiSi                   & 25                                          & 5.9147                                                                                    & \multicolumn{1}{c}{1.00}               & \multicolumn{1}{c}{0.20}  & \multicolumn{1}{c}{1.06}   & \multicolumn{1}{c}{-0.16} & -0.02 \\ 
			RuCrTiSi                   & 22                                          & 5.9792                                                                                    & \multicolumn{1}{c}{-1.99}              & \multicolumn{1}{c}{-0.18} & \multicolumn{1}{c}{-1.84}  & \multicolumn{1}{c}{0.18}  & 0.03  \\ 
			RuCrZrGa                   & 21                                          & 6.3195                                                                                    & \multicolumn{1}{c}{3.00}               & \multicolumn{1}{c}{0.08}  & \multicolumn{1}{c}{2.88}   & \multicolumn{1}{c}{-0.09} & -0.04 \\
		\end{tabular}
	\end{ruledtabular}
\end{table} 
\par In the case of the XMnCrZ compounds, an anomaly occurs. The so-called empirical \textquote{lighter-atom rule}, usually assumed to determine the sequence of those atoms in Heusler compounds demand that the sequence of the atoms along the diagonal is X-Cr-Mn-Z corresponding to the Type-3 structure.  But Mn atom is an anomaly in the periodic table. As we move from Sc to Ni along the 3d transition metal atoms, the electronegativity increases, and thus the \textquote{lighter-atom rule} reflects the electronegativity argument, mentioned above, which demands that the less electronegative transition metal atom in the sequence of the atoms is found in between the two other transition-metal atoms. But Mn atom has an electronegativity close to Ti and is much smaller than its neighboring Cr. Thus electronegativity arguments request that the sequence of atoms is X-Mn-Cr-Z corresponding to the Type-I structure. As mentioned in the paragraph above all five compounds containing both Cr and Mn atoms prefer to crystallize in the Type-I lattice structure as expected from the electronegativity argument and the empirical "lighter-atom rule" breaks down. This choice of the MnCr-based compounds can be easily understood in terms of their spin magnetic moments in the two different lattice types.
\par In type-1 structure, the Mn atoms carry the usual magnetic moment ranging nearly from 2.6 to 3.2$\mu_B$ while Cr atoms have magnetic moments in the range -1.3 to -1.7$\mu_B$ resulting in the integer value of the total magnetic moment. When we move to the Type-3 structure, the total magnetic moment is again the integer value and half-metallicity is present. The problem is with the atomic spin magnetic moments. Again the X and Z atoms carry small magnetic moments but Cr carries magnetic moments in the range of ~1.5 to 2.4$\mu_B$ and Mn atoms have very small magnetic moments in all cases except NiMnCrAl where its value is -1.1045$\mu_B$, which is still a small value for Mn atoms. The atomic DOS of Mn in most magnetic compounds is characterized by a large splitting between the majority-occupied and the minority-unoccupied bands leading to large values of its spin magnetic moment. Thus this situation is unphysical and this is why it is less favorable than the Type-1 structure. Also, let us take an example of IrMnCrSi and compare the values of spin moments of Mn and Cr atoms. If the Manganese atom is to have the spin moment of -3$\mu_B$ in type-3 structure, (which absolute value is equal to the spin moment in type-1 structure) the Cr atoms should have the magnitude of 5$\mu_B$ to keep the half-metallicity intact, which is also unphysical.
\par  In Figure \ref{fig:dos2}(b,d), we present two compounds that prefer to crystallize in a type-1 structure. One can see large half-metallic gap on both compounds in the spin-down states making them a strong candidate of half-metal. When we go from IrMnCrGe to IrMnCrSi, one can notice a large exchange splitting in the latter case. It is because in compounds usually the 3s and 3p states of Si are closer to the Fermi level than the 4s and 4p states of Ge and hence p-d hybridization is significant though Ge and Si have the same number of valence electrons. The spin resolved DOS and bandstructure of NiMnCrAl (not presented here) suggests that the compound is nearly SGS since the valence band and conduction band touch the $\Gamma$ and X points at the Fermi-level in the spin-up channel whereas there is a gap in the spin-down channel. However, the $e_u$ states on the conduction band slightly touch the Fermi level at $\Gamma$ point preventing it from being the ideal candidate of SGS.

\par  The individual and total spin magnetic moments of the compounds which prefer to crystallize in type-3 structure can be seen in Table \ref{tab:table_elastic}. The absolute value of spin moments ranges from 0 to 6$\mu_B$. Here, Mn whenever present have an absolute spin magnetic moment of ~3-3.5$\mu_B$, which also supports our argument for the stability of type-1 structure for the five compounds discussed above. Other than Manganese, Chromium and Cobalt atoms have considerable individual magnetic moments in the compounds while the main-group elements barely contribute to the observed total magnetic moments. The total magnetic moments of all of the compounds is integer (or nearly integer) value as predicted by the Slater-Pauling rule; the exception is RhCoZrAl where the total spin magnetic moment has significantly deviated from the integer value. The lattice variation in the same family of compounds can be associated with the size of atomic radii. For example if we compare RhFeTiSi and RhFeTiGe, Ge has a larger atomic radius than Si and thus the lattice constant of RhMnTiGe is larger. For the same reason the lattice constant of RhFeTiSi is larger than the lattice constant of RhFeMnSi since Ti has a larger atomic radius than Mn.

\par Among twenty compounds that crystallize in type-3 structure, two of the compounds, IrFeZrAl and IrRuTiAl, can be categorized as non-magnetic semiconductors as the total value of the spin magnetic moment is zero and the gap is present on both spin channels. For these compounds, we make several tests to be sure that it is not converged to a local minimum state. The test is performed starting from an antiferromagnetic initial distribution of the atomic spin magnetic moments but we again converged to the same non-magnetic ground state. Since both have 24 valence electrons, in order to be a half-metallic magnet Fe or Ru should have antiparallel spin magnetic moments with Zr or Ti in order to cancel out or they both must have zero magnetic moments. Since, Zr and Ti are harder to magnetize, these compounds prefer to be a semiconductor.   

\section{Conclusion}    
In conclusion, we have presented a detailed investigation of 25 quaternary Heusler compounds containing 3\textit{d}, 4\textit{d} and 5\textit{d} elements using the Quantum Espresso package. We start by screening a large number of potential compounds using the Open Quantum Materials Database and extend our study to calculate electronic, magnetic, and mechanical properties of those compounds that are feasible to synthesize in the lab. The thermodynamic and mechanical stability is ensured through convex-hull distance and elastic constants calculation. Among 25 studied compounds, we have identified 21 half-metals (or nearly half-metals), 2 spin-gapless semiconductor and 2 non-magnetic semi-conductor. The Slater-Pauling rule is followed by all of the compounds. The meticulous calculation is performed to find out the possible crystallized structure of the compounds among three non-equivalent superstructure. Our finding suggests that 5 of the compounds among 25 prefer to crystallize in type-1 structure while the rest of the compounds crystallize in type-3 structure. We believe that our study will augment the interest in Quaternary Heusler compounds for spintronics applications, providing experimentalists a new avenue for the design and synthesis of novel half-metallic compounds. 

\section{Acknowledgment}   
Calculations were performed partly with the computational resources provided by the Kathmandu University Supercomputer Center established with the equipment donated by CERN and other parts of computations were performed on the Wake Forest University DEAC Cluster, a centrally 
managed resource with support provided in part by the University. R. Adhikari would like to acknowledge GPU grants from NVIDIA. 
\bibliographystyle{apsrev4-2}
\bibliography{mybibfile}

\providecommand{\noopsort}[1]{}\providecommand{\singleletter}[1]{#1}%
\begin{thebibliography}{63}%
\makeatletter
\providecommand \@ifxundefined [1]{%
 \@ifx{#1\undefined}
}%
\providecommand \@ifnum [1]{%
 \ifnum #1\expandafter \@firstoftwo
 \else \expandafter \@secondoftwo
 \fi
}%
\providecommand \@ifx [1]{%
 \ifx #1\expandafter \@firstoftwo
 \else \expandafter \@secondoftwo
 \fi
}%
\providecommand \natexlab [1]{#1}%
\providecommand \enquote  [1]{``#1''}%
\providecommand \bibnamefont  [1]{#1}%
\providecommand \bibfnamefont [1]{#1}%
\providecommand \citenamefont [1]{#1}%
\providecommand \href@noop [0]{\@secondoftwo}%
\providecommand \href [0]{\begingroup \@sanitize@url \@href}%
\providecommand \@href[1]{\@@startlink{#1}\@@href}%
\providecommand \@@href[1]{\endgroup#1\@@endlink}%
\providecommand \@sanitize@url [0]{\catcode `\\12\catcode `\$12\catcode
  `\&12\catcode `\#12\catcode `\^12\catcode `\_12\catcode `\%12\relax}%
\providecommand \@@startlink[1]{}%
\providecommand \@@endlink[0]{}%
\providecommand \url  [0]{\begingroup\@sanitize@url \@url }%
\providecommand \@url [1]{\endgroup\@href {#1}{\urlprefix }}%
\providecommand \urlprefix  [0]{URL }%
\providecommand \Eprint [0]{\href }%
\providecommand \doibase [0]{https://doi.org/}%
\providecommand \selectlanguage [0]{\@gobble}%
\providecommand \bibinfo  [0]{\@secondoftwo}%
\providecommand \bibfield  [0]{\@secondoftwo}%
\providecommand \translation [1]{[#1]}%
\providecommand \BibitemOpen [0]{}%
\providecommand \bibitemStop [0]{}%
\providecommand \bibitemNoStop [0]{.\EOS\space}%
\providecommand \EOS [0]{\spacefactor3000\relax}%
\providecommand \BibitemShut  [1]{\csname bibitem#1\endcsname}%
\let\auto@bib@innerbib\@empty
\bibitem [{\citenamefont {de~Groot}\ \emph {et~al.}(1983)\citenamefont
  {de~Groot}, \citenamefont {Mueller}, \citenamefont {Engen},\ and\
  \citenamefont {Buschow}}]{deGroot}%
  \BibitemOpen
  \bibfield  {author} {\bibinfo {author} {\bibfnamefont {R.~A.}\ \bibnamefont
  {de~Groot}}, \bibinfo {author} {\bibfnamefont {F.~M.}\ \bibnamefont
  {Mueller}}, \bibinfo {author} {\bibfnamefont {P.~G.~v.}\ \bibnamefont
  {Engen}},\ and\ \bibinfo {author} {\bibfnamefont {K.~H.~J.}\ \bibnamefont
  {Buschow}},\ }\href {https://doi.org/10.1103/PhysRevLett.50.2024} {\bibfield
  {journal} {\bibinfo  {journal} {Phys. Rev. Lett.}\ }\textbf {\bibinfo
  {volume} {50}},\ \bibinfo {pages} {2024} (\bibinfo {year}
  {1983})}\BibitemShut {NoStop}%
\bibitem [{\citenamefont {\ifmmode \check{Z}\else
  \v{Z}\fi{}uti\ifmmode~\acute{c}\else \'{c}\fi{}}\ \emph
  {et~al.}(2004)\citenamefont {\ifmmode \check{Z}\else
  \v{Z}\fi{}uti\ifmmode~\acute{c}\else \'{c}\fi{}}, \citenamefont {Fabian},\
  and\ \citenamefont {Das~Sarma}}]{spintronics}%
  \BibitemOpen
  \bibfield  {author} {\bibinfo {author} {\bibfnamefont {I.}~\bibnamefont
  {\ifmmode \check{Z}\else \v{Z}\fi{}uti\ifmmode~\acute{c}\else \'{c}\fi{}}},
  \bibinfo {author} {\bibfnamefont {J.}~\bibnamefont {Fabian}},\ and\ \bibinfo
  {author} {\bibfnamefont {S.}~\bibnamefont {Das~Sarma}},\ }\href
  {https://doi.org/10.1103/RevModPhys.76.323} {\bibfield  {journal} {\bibinfo
  {journal} {Rev. Mod. Phys.}\ }\textbf {\bibinfo {volume} {76}},\ \bibinfo
  {pages} {323} (\bibinfo {year} {2004})}\BibitemShut {NoStop}%
\bibitem [{\citenamefont {Hirohata}\ and\ \citenamefont
  {Takanashi}(2014)}]{Hirohata}%
  \BibitemOpen
  \bibfield  {author} {\bibinfo {author} {\bibfnamefont {A.}~\bibnamefont
  {Hirohata}}\ and\ \bibinfo {author} {\bibfnamefont {K.}~\bibnamefont
  {Takanashi}},\ }\href {https://doi.org/10.1088/0022-3727/47/19/193001}
  {\bibfield  {journal} {\bibinfo  {journal} {J. Phys. D : Appl. Phys.}\
  }\textbf {\bibinfo {volume} {47}},\ \bibinfo {pages} {193001} (\bibinfo
  {year} {2014})}\BibitemShut {NoStop}%
\bibitem [{\citenamefont {Felser}\ and\ \citenamefont
  {Fecher}(2013)}]{SpintronicsClaudia}%
  \BibitemOpen
  \bibinfo {editor} {\bibfnamefont {C.}~\bibnamefont {Felser}}\ and\ \bibinfo
  {editor} {\bibfnamefont {G.~H.}\ \bibnamefont {Fecher}},\ eds.,\ \href@noop
  {} {\emph {\bibinfo {title} {{Spintronics: From Materials to Devices}}}},\
  \bibinfo {edition} {1st}\ ed.\ (\bibinfo  {publisher} {Springer},\ \bibinfo
  {address} {Netherlands},\ \bibinfo {year} {2013})\BibitemShut {NoStop}%
\bibitem [{\citenamefont {Galanakis}\ and\ \citenamefont
  {Dederichs}(2005)}]{HMAlloys-Galanakis-lec}%
  \BibitemOpen
  \bibinfo {editor} {\bibfnamefont {I.}~\bibnamefont {Galanakis}}\ and\
  \bibinfo {editor} {\bibfnamefont {P.~H.}\ \bibnamefont {Dederichs}},\ eds.,\
  \href@noop {} {\emph {\bibinfo {title} {{Half-metallic Alloys: Fundamentals
  and Applications}}}},\ \bibinfo {edition} {1st}\ ed.,\ \bibinfo {series}
  {Lecture Notes in Physics}, Vol.\ \bibinfo {volume} {676}\ (\bibinfo
  {publisher} {Springer-Verlag},\ \bibinfo {address} {Berlin Heidelberg},\
  \bibinfo {year} {2005})\BibitemShut {NoStop}%
\bibitem [{\citenamefont {Kato}\ \emph {et~al.}(2004)\citenamefont {Kato},
  \citenamefont {Okuda}, \citenamefont {Okimoto}, \citenamefont {Tomioka},
  \citenamefont {Oikawa}, \citenamefont {Kamiyama},\ and\ \citenamefont
  {Tokura}}]{kato2004structural}%
  \BibitemOpen
  \bibfield  {author} {\bibinfo {author} {\bibfnamefont {H.}~\bibnamefont
  {Kato}}, \bibinfo {author} {\bibfnamefont {T.}~\bibnamefont {Okuda}},
  \bibinfo {author} {\bibfnamefont {Y.}~\bibnamefont {Okimoto}}, \bibinfo
  {author} {\bibfnamefont {Y.}~\bibnamefont {Tomioka}}, \bibinfo {author}
  {\bibfnamefont {K.}~\bibnamefont {Oikawa}}, \bibinfo {author} {\bibfnamefont
  {T.}~\bibnamefont {Kamiyama}},\ and\ \bibinfo {author} {\bibfnamefont
  {Y.}~\bibnamefont {Tokura}},\ }\href@noop {} {\bibfield  {journal} {\bibinfo
  {journal} {Phys. Rev. B}\ }\textbf {\bibinfo {volume} {69}},\ \bibinfo
  {pages} {184412} (\bibinfo {year} {2004})}\BibitemShut {NoStop}%
\bibitem [{\citenamefont {Stroppa}\ \emph {et~al.}(2003)\citenamefont
  {Stroppa}, \citenamefont {Picozzi}, \citenamefont {Continenza},\ and\
  \citenamefont {Freeman}}]{stroppa2003electronic}%
  \BibitemOpen
  \bibfield  {author} {\bibinfo {author} {\bibfnamefont {A.}~\bibnamefont
  {Stroppa}}, \bibinfo {author} {\bibfnamefont {S.}~\bibnamefont {Picozzi}},
  \bibinfo {author} {\bibfnamefont {A.}~\bibnamefont {Continenza}},\ and\
  \bibinfo {author} {\bibfnamefont {A.}~\bibnamefont {Freeman}},\ }\href@noop
  {} {\bibfield  {journal} {\bibinfo  {journal} {Phys. Rev. B}\ }\textbf
  {\bibinfo {volume} {68}},\ \bibinfo {pages} {155203} (\bibinfo {year}
  {2003})}\BibitemShut {NoStop}%
\bibitem [{\citenamefont {Akai}(1998)}]{akai1998ferromagnetism}%
  \BibitemOpen
  \bibfield  {author} {\bibinfo {author} {\bibfnamefont {H.}~\bibnamefont
  {Akai}},\ }\href@noop {} {\bibfield  {journal} {\bibinfo  {journal} {Phys.
  Rev. Lett.}\ }\textbf {\bibinfo {volume} {81}},\ \bibinfo {pages} {3002}
  (\bibinfo {year} {1998})}\BibitemShut {NoStop}%
\bibitem [{\citenamefont {Soulen~Jr}\ \emph {et~al.}(1998)\citenamefont
  {Soulen~Jr}, \citenamefont {Byers}, \citenamefont {Osofsky}, \citenamefont
  {Nadgorny}, \citenamefont {Ambrose}, \citenamefont {Cheng}, \citenamefont
  {Broussard}, \citenamefont {Tanaka}, \citenamefont {Nowak}, \citenamefont
  {Moodera} \emph {et~al.}}]{soulen1998measuring}%
  \BibitemOpen
  \bibfield  {author} {\bibinfo {author} {\bibfnamefont {R.}~\bibnamefont
  {Soulen~Jr}}, \bibinfo {author} {\bibfnamefont {J.}~\bibnamefont {Byers}},
  \bibinfo {author} {\bibfnamefont {M.}~\bibnamefont {Osofsky}}, \bibinfo
  {author} {\bibfnamefont {B.}~\bibnamefont {Nadgorny}}, \bibinfo {author}
  {\bibfnamefont {T.}~\bibnamefont {Ambrose}}, \bibinfo {author} {\bibfnamefont
  {S.}~\bibnamefont {Cheng}}, \bibinfo {author} {\bibfnamefont {P.~R.}\
  \bibnamefont {Broussard}}, \bibinfo {author} {\bibfnamefont {C.}~\bibnamefont
  {Tanaka}}, \bibinfo {author} {\bibfnamefont {J.}~\bibnamefont {Nowak}},
  \bibinfo {author} {\bibfnamefont {J.}~\bibnamefont {Moodera}}, \emph
  {et~al.},\ }\href@noop {} {\bibfield  {journal} {\bibinfo  {journal}
  {Science}\ }\textbf {\bibinfo {volume} {282}},\ \bibinfo {pages} {85}
  (\bibinfo {year} {1998})}\BibitemShut {NoStop}%
\bibitem [{\citenamefont {Galanakis}\ and\ \citenamefont
  {Mavropoulos}(2003)}]{galanakis2003zinc}%
  \BibitemOpen
  \bibfield  {author} {\bibinfo {author} {\bibfnamefont {I.}~\bibnamefont
  {Galanakis}}\ and\ \bibinfo {author} {\bibfnamefont {P.}~\bibnamefont
  {Mavropoulos}},\ }\href@noop {} {\bibfield  {journal} {\bibinfo  {journal}
  {Phys. Rev. B}\ }\textbf {\bibinfo {volume} {67}},\ \bibinfo {pages} {104417}
  (\bibinfo {year} {2003})}\BibitemShut {NoStop}%
\bibitem [{\citenamefont {Galanakis}\ \emph {et~al.}(2002)\citenamefont
  {Galanakis}, \citenamefont {Dederichs},\ and\ \citenamefont
  {Papanikolaou}}]{Galanakis2002}%
  \BibitemOpen
  \bibfield  {author} {\bibinfo {author} {\bibfnamefont {I.}~\bibnamefont
  {Galanakis}}, \bibinfo {author} {\bibfnamefont {P.~H.}\ \bibnamefont
  {Dederichs}},\ and\ \bibinfo {author} {\bibfnamefont {N.}~\bibnamefont
  {Papanikolaou}},\ }\href {https://doi.org/10.1103/PhysRevB.66.174429}
  {\bibfield  {journal} {\bibinfo  {journal} {Phys. Rev. B}\ }\textbf {\bibinfo
  {volume} {66}},\ \bibinfo {pages} {174429} (\bibinfo {year}
  {2002})}\BibitemShut {NoStop}%
\bibitem [{\citenamefont {Skaftouros}\ \emph {et~al.}(2013)\citenamefont
  {Skaftouros}, \citenamefont {Özdoğan}, \citenamefont {Şaşıoğlu},\ and\
  \citenamefont {Galanakis}}]{Galanakisinverse}%
  \BibitemOpen
  \bibfield  {author} {\bibinfo {author} {\bibfnamefont {S.}~\bibnamefont
  {Skaftouros}}, \bibinfo {author} {\bibfnamefont {K.}~\bibnamefont
  {Özdoğan}}, \bibinfo {author} {\bibfnamefont {E.}~\bibnamefont
  {Şaşıoğlu}},\ and\ \bibinfo {author} {\bibfnamefont {I.}~\bibnamefont
  {Galanakis}},\ }\href {https://doi.org/10.1103/PhysRevB.87.024420} {\bibfield
   {journal} {\bibinfo  {journal} {Phys. Rev. B}\ }\textbf {\bibinfo {volume}
  {87}},\ \bibinfo {pages} {024420} (\bibinfo {year} {2013})}\BibitemShut
  {NoStop}%
\bibitem [{\citenamefont {{\"O}zdo{\u{g}}an}\ \emph {et~al.}(2013)\citenamefont
  {{\"O}zdo{\u{g}}an}, \citenamefont {{\c{S}}a{\c{s}}{\i}o{\u{g}}lu},\ and\
  \citenamefont {Galanakis}}]{ozdougan2013slater}%
  \BibitemOpen
  \bibfield  {author} {\bibinfo {author} {\bibfnamefont {K.}~\bibnamefont
  {{\"O}zdo{\u{g}}an}}, \bibinfo {author} {\bibfnamefont {E.}~\bibnamefont
  {{\c{S}}a{\c{s}}{\i}o{\u{g}}lu}},\ and\ \bibinfo {author} {\bibfnamefont
  {I.}~\bibnamefont {Galanakis}},\ }\href@noop {} {\bibfield  {journal}
  {\bibinfo  {journal} {J. Appl. Phys.}\ }\textbf {\bibinfo {volume} {113}},\
  \bibinfo {pages} {193903} (\bibinfo {year} {2013})}\BibitemShut {NoStop}%
\bibitem [{\citenamefont {Wurmehl}\ \emph {et~al.}(2005)\citenamefont
  {Wurmehl}, \citenamefont {Fecher}, \citenamefont {Kandpal}, \citenamefont
  {Ksenofontov}, \citenamefont {Felser}, \citenamefont {Lin},\ and\
  \citenamefont {Morais}}]{wurmehl2005geometric}%
  \BibitemOpen
  \bibfield  {author} {\bibinfo {author} {\bibfnamefont {S.}~\bibnamefont
  {Wurmehl}}, \bibinfo {author} {\bibfnamefont {G.~H.}\ \bibnamefont {Fecher}},
  \bibinfo {author} {\bibfnamefont {H.~C.}\ \bibnamefont {Kandpal}}, \bibinfo
  {author} {\bibfnamefont {V.}~\bibnamefont {Ksenofontov}}, \bibinfo {author}
  {\bibfnamefont {C.}~\bibnamefont {Felser}}, \bibinfo {author} {\bibfnamefont
  {H.-J.}\ \bibnamefont {Lin}},\ and\ \bibinfo {author} {\bibfnamefont
  {J.}~\bibnamefont {Morais}},\ }\href@noop {} {\bibfield  {journal} {\bibinfo
  {journal} {Phys. Rev. B}\ }\textbf {\bibinfo {volume} {72}},\ \bibinfo
  {pages} {184434} (\bibinfo {year} {2005})}\BibitemShut {NoStop}%
\bibitem [{\citenamefont {Felser}\ and\ \citenamefont
  {Hirohata}(2005)}]{HeuslerPropandGrowth}%
  \BibitemOpen
  \bibinfo {editor} {\bibfnamefont {C.}~\bibnamefont {Felser}}\ and\ \bibinfo
  {editor} {\bibfnamefont {A.}~\bibnamefont {Hirohata}},\ eds.,\ \href@noop {}
  {\emph {\bibinfo {title} {{Heusler Alloys: Properties, Growth,
  Applications}}}},\ \bibinfo {edition} {1st}\ ed.,\ \bibinfo {series}
  {Springer Series in Materials Science}, Vol.\ \bibinfo {volume} {222}\
  (\bibinfo  {publisher} {Springer International Publishing},\ \bibinfo
  {address} {Switzerland},\ \bibinfo {year} {2005})\BibitemShut {NoStop}%
\bibitem [{\citenamefont {Fong}\ \emph {et~al.}(2013)\citenamefont {Fong},
  \citenamefont {Pask},\ and\ \citenamefont {Yang}}]{fong2013half}%
  \BibitemOpen
  \bibfield  {author} {\bibinfo {author} {\bibfnamefont {C.-y.}\ \bibnamefont
  {Fong}}, \bibinfo {author} {\bibfnamefont {J.~E.}\ \bibnamefont {Pask}},\
  and\ \bibinfo {author} {\bibfnamefont {L.~H.}\ \bibnamefont {Yang}},\
  }\href@noop {} {\emph {\bibinfo {title} {{Half-metallic materials and their
  properties}}}},\ Vol.~\bibinfo {volume} {2}\ (\bibinfo  {publisher} {World
  Scientific},\ \bibinfo {year} {2013})\BibitemShut {NoStop}%
\bibitem [{\citenamefont {Block}\ \emph {et~al.}(2003)\citenamefont {Block},
  \citenamefont {Felser}, \citenamefont {Jakob}, \citenamefont {Ensling},
  \citenamefont {M{\"u}hling}, \citenamefont {G{\"u}tlich},\ and\ \citenamefont
  {Cava}}]{block2003large}%
  \BibitemOpen
  \bibfield  {author} {\bibinfo {author} {\bibfnamefont {T.}~\bibnamefont
  {Block}}, \bibinfo {author} {\bibfnamefont {C.}~\bibnamefont {Felser}},
  \bibinfo {author} {\bibfnamefont {G.}~\bibnamefont {Jakob}}, \bibinfo
  {author} {\bibfnamefont {J.}~\bibnamefont {Ensling}}, \bibinfo {author}
  {\bibfnamefont {B.}~\bibnamefont {M{\"u}hling}}, \bibinfo {author}
  {\bibfnamefont {P.}~\bibnamefont {G{\"u}tlich}},\ and\ \bibinfo {author}
  {\bibfnamefont {R.}~\bibnamefont {Cava}},\ }\href@noop {} {\bibfield
  {journal} {\bibinfo  {journal} {J. Solid State Chem.}\ }\textbf {\bibinfo
  {volume} {176}},\ \bibinfo {pages} {646} (\bibinfo {year}
  {2003})}\BibitemShut {NoStop}%
\bibitem [{\citenamefont {Felser}\ \emph {et~al.}(2003)\citenamefont {Felser},
  \citenamefont {Heitkamp}, \citenamefont {Kronast}, \citenamefont {Schmitz},
  \citenamefont {Cramm}, \citenamefont {D{\"u}rr}, \citenamefont {Elmers},
  \citenamefont {Fecher}, \citenamefont {Wurmehl}, \citenamefont {Block} \emph
  {et~al.}}]{felser2003investigation}%
  \BibitemOpen
  \bibfield  {author} {\bibinfo {author} {\bibfnamefont {C.}~\bibnamefont
  {Felser}}, \bibinfo {author} {\bibfnamefont {B.}~\bibnamefont {Heitkamp}},
  \bibinfo {author} {\bibfnamefont {F.}~\bibnamefont {Kronast}}, \bibinfo
  {author} {\bibfnamefont {D.}~\bibnamefont {Schmitz}}, \bibinfo {author}
  {\bibfnamefont {S.}~\bibnamefont {Cramm}}, \bibinfo {author} {\bibfnamefont
  {H.}~\bibnamefont {D{\"u}rr}}, \bibinfo {author} {\bibfnamefont
  {H.}~\bibnamefont {Elmers}}, \bibinfo {author} {\bibfnamefont
  {G.}~\bibnamefont {Fecher}}, \bibinfo {author} {\bibfnamefont
  {S.}~\bibnamefont {Wurmehl}}, \bibinfo {author} {\bibfnamefont
  {T.}~\bibnamefont {Block}}, \emph {et~al.},\ }\href@noop {} {\bibfield
  {journal} {\bibinfo  {journal} {J. Phys.: Condens. Matter}\ }\textbf
  {\bibinfo {volume} {15}},\ \bibinfo {pages} {7019} (\bibinfo {year}
  {2003})}\BibitemShut {NoStop}%
\bibitem [{\citenamefont {Galanakis}(2004)}]{galanakis2004appearance}%
  \BibitemOpen
  \bibfield  {author} {\bibinfo {author} {\bibfnamefont {I.}~\bibnamefont
  {Galanakis}},\ }\href@noop {} {\bibfield  {journal} {\bibinfo  {journal} {J.
  Phys.: Condens. Matter}\ }\textbf {\bibinfo {volume} {16}},\ \bibinfo {pages}
  {3089} (\bibinfo {year} {2004})}\BibitemShut {NoStop}%
\bibitem [{\citenamefont {Dai}\ \emph {et~al.}(2009)\citenamefont {Dai},
  \citenamefont {Liu}, \citenamefont {Fecher}, \citenamefont {Felser},
  \citenamefont {Li},\ and\ \citenamefont {Liu}}]{dai2009new}%
  \BibitemOpen
  \bibfield  {author} {\bibinfo {author} {\bibfnamefont {X.}~\bibnamefont
  {Dai}}, \bibinfo {author} {\bibfnamefont {G.}~\bibnamefont {Liu}}, \bibinfo
  {author} {\bibfnamefont {G.~H.}\ \bibnamefont {Fecher}}, \bibinfo {author}
  {\bibfnamefont {C.}~\bibnamefont {Felser}}, \bibinfo {author} {\bibfnamefont
  {Y.}~\bibnamefont {Li}},\ and\ \bibinfo {author} {\bibfnamefont
  {H.}~\bibnamefont {Liu}},\ }\href@noop {} {\bibfield  {journal} {\bibinfo
  {journal} {J. Appl. Phys.}\ }\textbf {\bibinfo {volume} {105}},\ \bibinfo
  {pages} {07E901} (\bibinfo {year} {2009})}\BibitemShut {NoStop}%
\bibitem [{\citenamefont {Alijani}\ \emph
  {et~al.}(2011{\natexlab{a}})\citenamefont {Alijani}, \citenamefont {Ouardi},
  \citenamefont {Fecher}, \citenamefont {Winterlik}, \citenamefont {Naghavi},
  \citenamefont {Kozina}, \citenamefont {Stryganyuk}, \citenamefont {Felser},
  \citenamefont {Ikenaga}, \citenamefont {Yamashita} \emph
  {et~al.}}]{alijani2011electronic}%
  \BibitemOpen
  \bibfield  {author} {\bibinfo {author} {\bibfnamefont {V.}~\bibnamefont
  {Alijani}}, \bibinfo {author} {\bibfnamefont {S.}~\bibnamefont {Ouardi}},
  \bibinfo {author} {\bibfnamefont {G.~H.}\ \bibnamefont {Fecher}}, \bibinfo
  {author} {\bibfnamefont {J.}~\bibnamefont {Winterlik}}, \bibinfo {author}
  {\bibfnamefont {S.~S.}\ \bibnamefont {Naghavi}}, \bibinfo {author}
  {\bibfnamefont {X.}~\bibnamefont {Kozina}}, \bibinfo {author} {\bibfnamefont
  {G.}~\bibnamefont {Stryganyuk}}, \bibinfo {author} {\bibfnamefont
  {C.}~\bibnamefont {Felser}}, \bibinfo {author} {\bibfnamefont
  {E.}~\bibnamefont {Ikenaga}}, \bibinfo {author} {\bibfnamefont
  {Y.}~\bibnamefont {Yamashita}}, \emph {et~al.},\ }\href@noop {} {\bibfield
  {journal} {\bibinfo  {journal} {Phys. Rev. B}\ }\textbf {\bibinfo {volume}
  {84}},\ \bibinfo {pages} {224416} (\bibinfo {year}
  {2011}{\natexlab{a}})}\BibitemShut {NoStop}%
\bibitem [{\citenamefont {Alijani}\ \emph
  {et~al.}(2011{\natexlab{b}})\citenamefont {Alijani}, \citenamefont
  {Winterlik}, \citenamefont {Fecher}, \citenamefont {Naghavi},\ and\
  \citenamefont {Felser}}]{alijani2011quaternary}%
  \BibitemOpen
  \bibfield  {author} {\bibinfo {author} {\bibfnamefont {V.}~\bibnamefont
  {Alijani}}, \bibinfo {author} {\bibfnamefont {J.}~\bibnamefont {Winterlik}},
  \bibinfo {author} {\bibfnamefont {G.~H.}\ \bibnamefont {Fecher}}, \bibinfo
  {author} {\bibfnamefont {S.~S.}\ \bibnamefont {Naghavi}},\ and\ \bibinfo
  {author} {\bibfnamefont {C.}~\bibnamefont {Felser}},\ }\href@noop {}
  {\bibfield  {journal} {\bibinfo  {journal} {Phys. Rev. B}\ }\textbf {\bibinfo
  {volume} {83}},\ \bibinfo {pages} {184428} (\bibinfo {year}
  {2011}{\natexlab{b}})}\BibitemShut {NoStop}%
\bibitem [{\citenamefont {Gao}\ \emph {et~al.}(2013)\citenamefont {Gao},
  \citenamefont {Hu}, \citenamefont {Yao}, \citenamefont {Luo},\ and\
  \citenamefont {Liu}}]{gao2013large}%
  \BibitemOpen
  \bibfield  {author} {\bibinfo {author} {\bibfnamefont {G.}~\bibnamefont
  {Gao}}, \bibinfo {author} {\bibfnamefont {L.}~\bibnamefont {Hu}}, \bibinfo
  {author} {\bibfnamefont {K.}~\bibnamefont {Yao}}, \bibinfo {author}
  {\bibfnamefont {B.}~\bibnamefont {Luo}},\ and\ \bibinfo {author}
  {\bibfnamefont {N.}~\bibnamefont {Liu}},\ }\href@noop {} {\bibfield
  {journal} {\bibinfo  {journal} {J. Alloys Cmpd.}\ }\textbf {\bibinfo {volume}
  {551}},\ \bibinfo {pages} {539} (\bibinfo {year} {2013})}\BibitemShut
  {NoStop}%
\bibitem [{\citenamefont {Gao}\ \emph {et~al.}(2019)\citenamefont {Gao},
  \citenamefont {Opahle},\ and\ \citenamefont {Zhang}}]{gao2019high}%
  \BibitemOpen
  \bibfield  {author} {\bibinfo {author} {\bibfnamefont {Q.}~\bibnamefont
  {Gao}}, \bibinfo {author} {\bibfnamefont {I.}~\bibnamefont {Opahle}},\ and\
  \bibinfo {author} {\bibfnamefont {H.}~\bibnamefont {Zhang}},\ }\href@noop {}
  {\bibfield  {journal} {\bibinfo  {journal} {Phys. Rev. Mater.}\ }\textbf
  {\bibinfo {volume} {3}},\ \bibinfo {pages} {024410} (\bibinfo {year}
  {2019})}\BibitemShut {NoStop}%
\bibitem [{\citenamefont {Aull}\ \emph {et~al.}(2019)\citenamefont {Aull},
  \citenamefont {{\c{S}}a{\c{s}}{\i}o{\u{g}}lu}, \citenamefont {Maznichenko},
  \citenamefont {Ostanin}, \citenamefont {Ernst}, \citenamefont {Mertig},\ and\
  \citenamefont {Galanakis}}]{aull2019ab}%
  \BibitemOpen
  \bibfield  {author} {\bibinfo {author} {\bibfnamefont {T.}~\bibnamefont
  {Aull}}, \bibinfo {author} {\bibfnamefont {E.}~\bibnamefont
  {{\c{S}}a{\c{s}}{\i}o{\u{g}}lu}}, \bibinfo {author} {\bibfnamefont
  {I.}~\bibnamefont {Maznichenko}}, \bibinfo {author} {\bibfnamefont
  {S.}~\bibnamefont {Ostanin}}, \bibinfo {author} {\bibfnamefont
  {A.}~\bibnamefont {Ernst}}, \bibinfo {author} {\bibfnamefont
  {I.}~\bibnamefont {Mertig}},\ and\ \bibinfo {author} {\bibfnamefont
  {I.}~\bibnamefont {Galanakis}},\ }\href@noop {} {\bibfield  {journal}
  {\bibinfo  {journal} {Phys. Rev. Mater.}\ }\textbf {\bibinfo {volume} {3}},\
  \bibinfo {pages} {124415} (\bibinfo {year} {2019})}\BibitemShut {NoStop}%
\bibitem [{\citenamefont {Saal}\ \emph {et~al.}(2013)\citenamefont {Saal},
  \citenamefont {Kirklin}, \citenamefont {Aykol}, \citenamefont {Meredig},\
  and\ \citenamefont {Wolverton}}]{OQMD1}%
  \BibitemOpen
  \bibfield  {author} {\bibinfo {author} {\bibfnamefont {J.~E.}\ \bibnamefont
  {Saal}}, \bibinfo {author} {\bibfnamefont {S.}~\bibnamefont {Kirklin}},
  \bibinfo {author} {\bibfnamefont {M.}~\bibnamefont {Aykol}}, \bibinfo
  {author} {\bibfnamefont {B.}~\bibnamefont {Meredig}},\ and\ \bibinfo {author}
  {\bibfnamefont {C.}~\bibnamefont {Wolverton}},\ }\href@noop {} {\bibfield
  {journal} {\bibinfo  {journal} {JOM}\ }\textbf {\bibinfo {volume} {65}},\
  \bibinfo {pages} {1501} (\bibinfo {year} {2013})}\BibitemShut {NoStop}%
\bibitem [{\citenamefont {Kirklin}\ \emph {et~al.}(2015)\citenamefont
  {Kirklin}, \citenamefont {Saal}, \citenamefont {Meredig}, \citenamefont
  {Thompson}, \citenamefont {Doak}, \citenamefont {Aykol}, \citenamefont
  {R{\"u}hl},\ and\ \citenamefont {Wolverton}}]{OQMD2}%
  \BibitemOpen
  \bibfield  {author} {\bibinfo {author} {\bibfnamefont {S.}~\bibnamefont
  {Kirklin}}, \bibinfo {author} {\bibfnamefont {J.~E.}\ \bibnamefont {Saal}},
  \bibinfo {author} {\bibfnamefont {B.}~\bibnamefont {Meredig}}, \bibinfo
  {author} {\bibfnamefont {A.}~\bibnamefont {Thompson}}, \bibinfo {author}
  {\bibfnamefont {J.~W.}\ \bibnamefont {Doak}}, \bibinfo {author}
  {\bibfnamefont {M.}~\bibnamefont {Aykol}}, \bibinfo {author} {\bibfnamefont
  {S.}~\bibnamefont {R{\"u}hl}},\ and\ \bibinfo {author} {\bibfnamefont
  {C.}~\bibnamefont {Wolverton}},\ }\href@noop {} {\bibfield  {journal}
  {\bibinfo  {journal} {Npj Comput. Mater.}\ }\textbf {\bibinfo {volume} {1}},\
  \bibinfo {pages} {1} (\bibinfo {year} {2015})}\BibitemShut {NoStop}%
\bibitem [{\citenamefont {Dhakal}\ \emph {et~al.}(2021)\citenamefont {Dhakal},
  \citenamefont {Nepal}, \citenamefont {Galanakis}, \citenamefont {Adhikari},\
  and\ \citenamefont {Kaphle}}]{dhakal2021prediction}%
  \BibitemOpen
  \bibfield  {author} {\bibinfo {author} {\bibfnamefont {R.}~\bibnamefont
  {Dhakal}}, \bibinfo {author} {\bibfnamefont {S.}~\bibnamefont {Nepal}},
  \bibinfo {author} {\bibfnamefont {I.}~\bibnamefont {Galanakis}}, \bibinfo
  {author} {\bibfnamefont {R.~P.}\ \bibnamefont {Adhikari}},\ and\ \bibinfo
  {author} {\bibfnamefont {G.~C.}\ \bibnamefont {Kaphle}},\ }\href@noop {}
  {\bibfield  {journal} {\bibinfo  {journal} {J. Alloys Compd.}\ }\textbf
  {\bibinfo {volume} {882}},\ \bibinfo {pages} {160500} (\bibinfo {year}
  {2021})}\BibitemShut {NoStop}%
\bibitem [{\citenamefont {Bainsla}\ \emph {et~al.}(2015)\citenamefont
  {Bainsla}, \citenamefont {Raja}, \citenamefont {Nigam},\ and\ \citenamefont
  {Suresh}}]{bainsla2015corufex}%
  \BibitemOpen
  \bibfield  {author} {\bibinfo {author} {\bibfnamefont {L.}~\bibnamefont
  {Bainsla}}, \bibinfo {author} {\bibfnamefont {M.~M.}\ \bibnamefont {Raja}},
  \bibinfo {author} {\bibfnamefont {A.}~\bibnamefont {Nigam}},\ and\ \bibinfo
  {author} {\bibfnamefont {K.}~\bibnamefont {Suresh}},\ }\href@noop {}
  {\bibfield  {journal} {\bibinfo  {journal} {J. Alloys Cmpd.}\ }\textbf
  {\bibinfo {volume} {651}},\ \bibinfo {pages} {631} (\bibinfo {year}
  {2015})}\BibitemShut {NoStop}%
\bibitem [{\citenamefont {Wang}\ \emph
  {et~al.}(2017{\natexlab{a}})\citenamefont {Wang}, \citenamefont {Khachai},
  \citenamefont {Khenata}, \citenamefont {Yuan}, \citenamefont {Wang},
  \citenamefont {Wang}, \citenamefont {Bouhemadou}, \citenamefont {Hao},
  \citenamefont {Dai}, \citenamefont {Guo} \emph
  {et~al.}}]{wang2017structural}%
  \BibitemOpen
  \bibfield  {author} {\bibinfo {author} {\bibfnamefont {X.}~\bibnamefont
  {Wang}}, \bibinfo {author} {\bibfnamefont {H.}~\bibnamefont {Khachai}},
  \bibinfo {author} {\bibfnamefont {R.}~\bibnamefont {Khenata}}, \bibinfo
  {author} {\bibfnamefont {H.}~\bibnamefont {Yuan}}, \bibinfo {author}
  {\bibfnamefont {L.}~\bibnamefont {Wang}}, \bibinfo {author} {\bibfnamefont
  {W.}~\bibnamefont {Wang}}, \bibinfo {author} {\bibfnamefont {A.}~\bibnamefont
  {Bouhemadou}}, \bibinfo {author} {\bibfnamefont {L.}~\bibnamefont {Hao}},
  \bibinfo {author} {\bibfnamefont {X.}~\bibnamefont {Dai}}, \bibinfo {author}
  {\bibfnamefont {R.}~\bibnamefont {Guo}}, \emph {et~al.},\ }\href@noop {}
  {\bibfield  {journal} {\bibinfo  {journal} {Sci. Rep.}\ }\textbf {\bibinfo
  {volume} {7}},\ \bibinfo {pages} {1} (\bibinfo {year}
  {2017}{\natexlab{a}})}\BibitemShut {NoStop}%
\bibitem [{\citenamefont {Berri}\ \emph {et~al.}(2014)\citenamefont {Berri},
  \citenamefont {Ibrir}, \citenamefont {Maouche},\ and\ \citenamefont
  {Attallah}}]{berri2014robust}%
  \BibitemOpen
  \bibfield  {author} {\bibinfo {author} {\bibfnamefont {S.}~\bibnamefont
  {Berri}}, \bibinfo {author} {\bibfnamefont {M.}~\bibnamefont {Ibrir}},
  \bibinfo {author} {\bibfnamefont {D.}~\bibnamefont {Maouche}},\ and\ \bibinfo
  {author} {\bibfnamefont {M.}~\bibnamefont {Attallah}},\ }\href@noop {}
  {\bibfield  {journal} {\bibinfo  {journal} {Comput. Condens. Matter}\
  }\textbf {\bibinfo {volume} {1}},\ \bibinfo {pages} {26} (\bibinfo {year}
  {2014})}\BibitemShut {NoStop}%
\bibitem [{\citenamefont {Kundu}\ \emph {et~al.}(2017)\citenamefont {Kundu},
  \citenamefont {Ghosh}, \citenamefont {Banerjee}, \citenamefont {Ghosh},\ and\
  \citenamefont {Sanyal}}]{kundu2017new}%
  \BibitemOpen
  \bibfield  {author} {\bibinfo {author} {\bibfnamefont {A.}~\bibnamefont
  {Kundu}}, \bibinfo {author} {\bibfnamefont {S.}~\bibnamefont {Ghosh}},
  \bibinfo {author} {\bibfnamefont {R.}~\bibnamefont {Banerjee}}, \bibinfo
  {author} {\bibfnamefont {S.}~\bibnamefont {Ghosh}},\ and\ \bibinfo {author}
  {\bibfnamefont {B.}~\bibnamefont {Sanyal}},\ }\href@noop {} {\bibfield
  {journal} {\bibinfo  {journal} {Sci. Rep.}\ }\textbf {\bibinfo {volume}
  {7}},\ \bibinfo {pages} {1} (\bibinfo {year} {2017})}\BibitemShut {NoStop}%
\bibitem [{\citenamefont {Labar}\ \emph {et~al.}(2021)\citenamefont {Labar},
  \citenamefont {Shankar}, \citenamefont {Ram}, \citenamefont {Laref},\ and\
  \citenamefont {Sharma}}]{labar2021novel}%
  \BibitemOpen
  \bibfield  {author} {\bibinfo {author} {\bibfnamefont {K.}~\bibnamefont
  {Labar}}, \bibinfo {author} {\bibfnamefont {A.}~\bibnamefont {Shankar}},
  \bibinfo {author} {\bibfnamefont {M.}~\bibnamefont {Ram}}, \bibinfo {author}
  {\bibfnamefont {A.}~\bibnamefont {Laref}},\ and\ \bibinfo {author}
  {\bibfnamefont {R.}~\bibnamefont {Sharma}},\ }\href@noop {} {\bibfield
  {journal} {\bibinfo  {journal} {J. Phys. Chem. Solids}\ }\textbf {\bibinfo
  {volume} {156}},\ \bibinfo {pages} {110119} (\bibinfo {year}
  {2021})}\BibitemShut {NoStop}%
\bibitem [{\citenamefont {Wang}\ \emph
  {et~al.}(2017{\natexlab{b}})\citenamefont {Wang}, \citenamefont {Cheng},
  \citenamefont {Guo}, \citenamefont {Wang}, \citenamefont {Rozale},
  \citenamefont {Wang}, \citenamefont {Yu},\ and\ \citenamefont
  {Liu}}]{wang2017first}%
  \BibitemOpen
  \bibfield  {author} {\bibinfo {author} {\bibfnamefont {X.}~\bibnamefont
  {Wang}}, \bibinfo {author} {\bibfnamefont {Z.}~\bibnamefont {Cheng}},
  \bibinfo {author} {\bibfnamefont {R.}~\bibnamefont {Guo}}, \bibinfo {author}
  {\bibfnamefont {J.}~\bibnamefont {Wang}}, \bibinfo {author} {\bibfnamefont
  {H.}~\bibnamefont {Rozale}}, \bibinfo {author} {\bibfnamefont
  {L.}~\bibnamefont {Wang}}, \bibinfo {author} {\bibfnamefont {Z.}~\bibnamefont
  {Yu}},\ and\ \bibinfo {author} {\bibfnamefont {G.}~\bibnamefont {Liu}},\
  }\href@noop {} {\bibfield  {journal} {\bibinfo  {journal} {Mater. Chem.
  Phys.}\ }\textbf {\bibinfo {volume} {193}},\ \bibinfo {pages} {99} (\bibinfo
  {year} {2017}{\natexlab{b}})}\BibitemShut {NoStop}%
\bibitem [{\citenamefont {Forozani}\ \emph {et~al.}(2020)\citenamefont
  {Forozani}, \citenamefont {Abadi}, \citenamefont {Baizaee},\ and\
  \citenamefont {Gharaati}}]{forozani2020structural}%
  \BibitemOpen
  \bibfield  {author} {\bibinfo {author} {\bibfnamefont {G.}~\bibnamefont
  {Forozani}}, \bibinfo {author} {\bibfnamefont {A.~A.~M.}\ \bibnamefont
  {Abadi}}, \bibinfo {author} {\bibfnamefont {S.~M.}\ \bibnamefont {Baizaee}},\
  and\ \bibinfo {author} {\bibfnamefont {A.}~\bibnamefont {Gharaati}},\
  }\href@noop {} {\bibfield  {journal} {\bibinfo  {journal} {J. Alloys Cmpd.}\
  }\textbf {\bibinfo {volume} {815}},\ \bibinfo {pages} {152449} (\bibinfo
  {year} {2020})}\BibitemShut {NoStop}%
\bibitem [{\citenamefont {Chinnadurai}\ and\ \citenamefont
  {Natesan}(2021)}]{chinnadurai2021first}%
  \BibitemOpen
  \bibfield  {author} {\bibinfo {author} {\bibfnamefont {K.}~\bibnamefont
  {Chinnadurai}}\ and\ \bibinfo {author} {\bibfnamefont {B.}~\bibnamefont
  {Natesan}},\ }\href@noop {} {\bibfield  {journal} {\bibinfo  {journal}
  {Comput. Mater. Sci.}\ }\textbf {\bibinfo {volume} {188}},\ \bibinfo {pages}
  {110116} (\bibinfo {year} {2021})}\BibitemShut {NoStop}%
\bibitem [{\citenamefont {Seh}\ and\ \citenamefont
  {Gupta}(2019)}]{pughcriticalelastic_Seh}%
  \BibitemOpen
  \bibfield  {author} {\bibinfo {author} {\bibfnamefont {A.~Q.}\ \bibnamefont
  {Seh}}\ and\ \bibinfo {author} {\bibfnamefont {D.~C.}\ \bibnamefont
  {Gupta}},\ }\href {https://doi.org/https://doi.org/10.1002/er.4853}
  {\bibfield  {journal} {\bibinfo  {journal} {Int. J. Energy Res.}\ }\textbf
  {\bibinfo {volume} {43}},\ \bibinfo {pages} {8864} (\bibinfo {year}
  {2019})},\ \Eprint
  {https://arxiv.org/abs/https://onlinelibrary.wiley.com/doi/pdf/10.1002/er.4853}
  {https://onlinelibrary.wiley.com/doi/pdf/10.1002/er.4853} \BibitemShut
  {NoStop}%
\bibitem [{\citenamefont {Li}\ \emph {et~al.}(2021)\citenamefont {Li},
  \citenamefont {Zhu}, \citenamefont {Paudel}, \citenamefont {Huang},\ and\
  \citenamefont {Zhou}}]{recent1}%
  \BibitemOpen
  \bibfield  {author} {\bibinfo {author} {\bibfnamefont {Y.}~\bibnamefont
  {Li}}, \bibinfo {author} {\bibfnamefont {J.}~\bibnamefont {Zhu}}, \bibinfo
  {author} {\bibfnamefont {R.}~\bibnamefont {Paudel}}, \bibinfo {author}
  {\bibfnamefont {J.}~\bibnamefont {Huang}},\ and\ \bibinfo {author}
  {\bibfnamefont {F.}~\bibnamefont {Zhou}},\ }\href@noop {} {\bibfield
  {journal} {\bibinfo  {journal} {Vacuum}\ }\textbf {\bibinfo {volume} {192}},\
  \bibinfo {pages} {110418} (\bibinfo {year} {2021})}\BibitemShut {NoStop}%
\bibitem [{\citenamefont {Prakash}\ \emph {et~al.}(2022)\citenamefont
  {Prakash}, \citenamefont {Suganya},\ and\ \citenamefont {Kalpana}}]{recent2}%
  \BibitemOpen
  \bibfield  {author} {\bibinfo {author} {\bibfnamefont {R.}~\bibnamefont
  {Prakash}}, \bibinfo {author} {\bibfnamefont {G.}~\bibnamefont {Suganya}},\
  and\ \bibinfo {author} {\bibfnamefont {G.}~\bibnamefont {Kalpana}},\
  }\href@noop {} {\bibfield  {journal} {\bibinfo  {journal} {AIP Advances}\
  }\textbf {\bibinfo {volume} {12}},\ \bibinfo {pages} {055223} (\bibinfo
  {year} {2022})}\BibitemShut {NoStop}%
\bibitem [{\citenamefont {Guo}\ \emph {et~al.}(2018)\citenamefont {Guo},
  \citenamefont {Ni}, \citenamefont {Liang},\ and\ \citenamefont
  {Luo}}]{recent3}%
  \BibitemOpen
  \bibfield  {author} {\bibinfo {author} {\bibfnamefont {X.}~\bibnamefont
  {Guo}}, \bibinfo {author} {\bibfnamefont {Z.}~\bibnamefont {Ni}}, \bibinfo
  {author} {\bibfnamefont {Z.}~\bibnamefont {Liang}},\ and\ \bibinfo {author}
  {\bibfnamefont {H.}~\bibnamefont {Luo}},\ }\href@noop {} {\bibfield
  {journal} {\bibinfo  {journal} {Comput. Mater. Sci.}\ }\textbf {\bibinfo
  {volume} {154}},\ \bibinfo {pages} {442} (\bibinfo {year}
  {2018})}\BibitemShut {NoStop}%
\bibitem [{\citenamefont {Dergal}\ \emph {et~al.}(2016)\citenamefont {Dergal},
  \citenamefont {Doumi}, \citenamefont {Mokaddem}, \citenamefont {Mamoun},\
  and\ \citenamefont {Merad}}]{recent4}%
  \BibitemOpen
  \bibfield  {author} {\bibinfo {author} {\bibfnamefont {S.}~\bibnamefont
  {Dergal}}, \bibinfo {author} {\bibfnamefont {B.}~\bibnamefont {Doumi}},
  \bibinfo {author} {\bibfnamefont {A.}~\bibnamefont {Mokaddem}}, \bibinfo
  {author} {\bibfnamefont {S.}~\bibnamefont {Mamoun}},\ and\ \bibinfo {author}
  {\bibfnamefont {A.~E.}\ \bibnamefont {Merad}},\ }\href@noop {} {\bibfield
  {journal} {\bibinfo  {journal} {J. Supercond. Nov. Magn.}\ }\textbf {\bibinfo
  {volume} {29}},\ \bibinfo {pages} {2953} (\bibinfo {year}
  {2016})}\BibitemShut {NoStop}%
\bibitem [{\citenamefont {Pandey}\ and\ \citenamefont
  {Kaphle}(2021)}]{recent5}%
  \BibitemOpen
  \bibfield  {author} {\bibinfo {author} {\bibfnamefont {B.}~\bibnamefont
  {Pandey}}\ and\ \bibinfo {author} {\bibfnamefont {G.~C.}\ \bibnamefont
  {Kaphle}},\ }\href@noop {} {\bibfield  {journal} {\bibinfo  {journal} {Mater.
  Today:Proc}\ }\textbf {\bibinfo {volume} {47}},\ \bibinfo {pages} {6481}
  (\bibinfo {year} {2021})}\BibitemShut {NoStop}%
\bibitem [{\citenamefont {Mohamedi}\ \emph {et~al.}(2016)\citenamefont
  {Mohamedi}, \citenamefont {Chahed}, \citenamefont {Amar}, \citenamefont
  {Rozale}, \citenamefont {Lakdja}, \citenamefont {Benhelal},\ and\
  \citenamefont {Sayede}}]{recent6}%
  \BibitemOpen
  \bibfield  {author} {\bibinfo {author} {\bibfnamefont {M.~W.}\ \bibnamefont
  {Mohamedi}}, \bibinfo {author} {\bibfnamefont {A.}~\bibnamefont {Chahed}},
  \bibinfo {author} {\bibfnamefont {A.}~\bibnamefont {Amar}}, \bibinfo {author}
  {\bibfnamefont {H.}~\bibnamefont {Rozale}}, \bibinfo {author} {\bibfnamefont
  {A.}~\bibnamefont {Lakdja}}, \bibinfo {author} {\bibfnamefont
  {O.}~\bibnamefont {Benhelal}},\ and\ \bibinfo {author} {\bibfnamefont
  {A.}~\bibnamefont {Sayede}},\ }\href@noop {} {\bibfield  {journal} {\bibinfo
  {journal} {Eur. Phys. J. B}\ }\textbf {\bibinfo {volume} {89}},\ \bibinfo
  {pages} {1} (\bibinfo {year} {2016})}\BibitemShut {NoStop}%
\bibitem [{\citenamefont {Xu}\ \emph {et~al.}(2013)\citenamefont {Xu},
  \citenamefont {Liu}, \citenamefont {Du}, \citenamefont {Li}, \citenamefont
  {Liu}, \citenamefont {Wang},\ and\ \citenamefont {Wu}}]{recent7}%
  \BibitemOpen
  \bibfield  {author} {\bibinfo {author} {\bibfnamefont {G.}~\bibnamefont
  {Xu}}, \bibinfo {author} {\bibfnamefont {E.}~\bibnamefont {Liu}}, \bibinfo
  {author} {\bibfnamefont {Y.}~\bibnamefont {Du}}, \bibinfo {author}
  {\bibfnamefont {G.}~\bibnamefont {Li}}, \bibinfo {author} {\bibfnamefont
  {G.}~\bibnamefont {Liu}}, \bibinfo {author} {\bibfnamefont {W.}~\bibnamefont
  {Wang}},\ and\ \bibinfo {author} {\bibfnamefont {G.}~\bibnamefont {Wu}},\
  }\href@noop {} {\bibfield  {journal} {\bibinfo  {journal} {Europhys. Lett.}\
  }\textbf {\bibinfo {volume} {102}},\ \bibinfo {pages} {17007} (\bibinfo
  {year} {2013})}\BibitemShut {NoStop}%
\bibitem [{\citenamefont {Giannozzi}\ \emph {et~al.}(2009)\citenamefont
  {Giannozzi}, \citenamefont {Baroni}, \citenamefont {Bonini}, \citenamefont
  {Calandra}, \citenamefont {Car},\ and\ \citenamefont {\textit{et
  al.}}}]{QE-2009}%
  \BibitemOpen
  \bibfield  {author} {\bibinfo {author} {\bibfnamefont {P.}~\bibnamefont
  {Giannozzi}}, \bibinfo {author} {\bibfnamefont {S.}~\bibnamefont {Baroni}},
  \bibinfo {author} {\bibfnamefont {N.}~\bibnamefont {Bonini}}, \bibinfo
  {author} {\bibfnamefont {M.}~\bibnamefont {Calandra}}, \bibinfo {author}
  {\bibfnamefont {R.}~\bibnamefont {Car}},\ and\ \bibinfo {author}
  {\bibnamefont {\textit{et al.}}},\ }\href {http://www.quantum-espresso.org}
  {\bibfield  {journal} {\bibinfo  {journal} {J. Phys.: Condens. Matter}\
  }\textbf {\bibinfo {volume} {21}},\ \bibinfo {pages} {395502 (19pp)}
  (\bibinfo {year} {2009})}\BibitemShut {NoStop}%
\bibitem [{\citenamefont {Giannozzi}\ \emph {et~al.}(2017)\citenamefont
  {Giannozzi}, \citenamefont {Andreussi}, \citenamefont {Brumme}, \citenamefont
  {Bunau}, \citenamefont {Nardelli},\ and\ \citenamefont {\textit{et
  al.}}}]{QE-2017}%
  \BibitemOpen
  \bibfield  {author} {\bibinfo {author} {\bibfnamefont {P.}~\bibnamefont
  {Giannozzi}}, \bibinfo {author} {\bibfnamefont {O.}~\bibnamefont
  {Andreussi}}, \bibinfo {author} {\bibfnamefont {T.}~\bibnamefont {Brumme}},
  \bibinfo {author} {\bibfnamefont {O.}~\bibnamefont {Bunau}}, \bibinfo
  {author} {\bibfnamefont {M.~B.}\ \bibnamefont {Nardelli}},\ and\ \bibinfo
  {author} {\bibnamefont {\textit{et al.}}},\ }\href
  {http://stacks.iop.org/0953-8984/29/i=46/a=465901} {\bibfield  {journal}
  {\bibinfo  {journal} {J. Phys.: Condens. Matter}\ }\textbf {\bibinfo {volume}
  {29}},\ \bibinfo {pages} {465901} (\bibinfo {year} {2017})}\BibitemShut
  {NoStop}%
\bibitem [{\citenamefont {Vanderbilt}(1990)}]{ultrasoft}%
  \BibitemOpen
  \bibfield  {author} {\bibinfo {author} {\bibfnamefont {D.}~\bibnamefont
  {Vanderbilt}},\ }\href {https://doi.org/10.1103/PhysRevB.41.7892} {\bibfield
  {journal} {\bibinfo  {journal} {Phys. Rev. B}\ }\textbf {\bibinfo {volume}
  {41}},\ \bibinfo {pages} {7892} (\bibinfo {year} {1990})}\BibitemShut
  {NoStop}%
\bibitem [{\citenamefont {Perdew}\ \emph {et~al.}(1996)\citenamefont {Perdew},
  \citenamefont {Burke},\ and\ \citenamefont
  {Ernzerhof}}]{perdew1996generalized}%
  \BibitemOpen
  \bibfield  {author} {\bibinfo {author} {\bibfnamefont {J.~P.}\ \bibnamefont
  {Perdew}}, \bibinfo {author} {\bibfnamefont {K.}~\bibnamefont {Burke}},\ and\
  \bibinfo {author} {\bibfnamefont {M.}~\bibnamefont {Ernzerhof}},\ }\href@noop
  {} {\bibfield  {journal} {\bibinfo  {journal} {Phys. Rev. Lett.}\ }\textbf
  {\bibinfo {volume} {77}},\ \bibinfo {pages} {3865} (\bibinfo {year}
  {1996})}\BibitemShut {NoStop}%
\bibitem [{\citenamefont {Bl\"ochl}\ \emph {et~al.}(1994)\citenamefont
  {Bl\"ochl}, \citenamefont {Jepsen},\ and\ \citenamefont
  {Andersen}}]{lineartetrahedra}%
  \BibitemOpen
  \bibfield  {author} {\bibinfo {author} {\bibfnamefont {P.~E.}\ \bibnamefont
  {Bl\"ochl}}, \bibinfo {author} {\bibfnamefont {O.}~\bibnamefont {Jepsen}},\
  and\ \bibinfo {author} {\bibfnamefont {O.~K.}\ \bibnamefont {Andersen}},\
  }\href {https://doi.org/10.1103/PhysRevB.49.16223} {\bibfield  {journal}
  {\bibinfo  {journal} {Phys. Rev. B}\ }\textbf {\bibinfo {volume} {49}},\
  \bibinfo {pages} {16223} (\bibinfo {year} {1994})}\BibitemShut {NoStop}%
\bibitem [{\citenamefont {Golesorkhtabar}\ \emph {et~al.}(2013)\citenamefont
  {Golesorkhtabar}, \citenamefont {Pavone}, \citenamefont {Spitaler},
  \citenamefont {Puschnig},\ and\ \citenamefont {Draxl}}]{elasticTool}%
  \BibitemOpen
  \bibfield  {author} {\bibinfo {author} {\bibfnamefont {R.}~\bibnamefont
  {Golesorkhtabar}}, \bibinfo {author} {\bibfnamefont {P.}~\bibnamefont
  {Pavone}}, \bibinfo {author} {\bibfnamefont {J.}~\bibnamefont {Spitaler}},
  \bibinfo {author} {\bibfnamefont {P.}~\bibnamefont {Puschnig}},\ and\
  \bibinfo {author} {\bibfnamefont {C.}~\bibnamefont {Draxl}},\ }\href
  {https://doi.org/https://doi.org/10.1016/j.cpc.2013.03.010} {\bibfield
  {journal} {\bibinfo  {journal} {Comput. Phys. Commun.}\ }\textbf {\bibinfo
  {volume} {184}},\ \bibinfo {pages} {1861 } (\bibinfo {year}
  {2013})}\BibitemShut {NoStop}%
\bibitem [{\citenamefont {Born}\ and\ \citenamefont
  {Huang}(1956)}]{born_huang_1956}%
  \BibitemOpen
  \bibfield  {author} {\bibinfo {author} {\bibfnamefont {M.}~\bibnamefont
  {Born}}\ and\ \bibinfo {author} {\bibfnamefont {K.}~\bibnamefont {Huang}},\
  }\href@noop {} {\emph {\bibinfo {title} {Dynamical theory of crystal
  lattices}}}\ (\bibinfo  {publisher} {Clarendon Press},\ \bibinfo {year}
  {1956})\BibitemShut {NoStop}%
\bibitem [{\citenamefont {Voigt}(1910)}]{voigt1910lehrbuch}%
  \BibitemOpen
  \bibfield  {author} {\bibinfo {author} {\bibfnamefont {W.}~\bibnamefont
  {Voigt}},\ }\href@noop {} {\emph {\bibinfo {title} {Lehrbuch der
  kristallphysik:(mit ausschluss der kristalloptik)}}},\ Vol.~\bibinfo {volume}
  {34}\ (\bibinfo  {publisher} {BG Teubner},\ \bibinfo {year}
  {1910})\BibitemShut {NoStop}%
\bibitem [{\citenamefont {Reuss}(1929)}]{ruess1929}%
  \BibitemOpen
  \bibfield  {author} {\bibinfo {author} {\bibfnamefont {A.}~\bibnamefont
  {Reuss}},\ }\href {https://doi.org/https://doi.org/10.1002/zamm.19290090104}
  {\bibfield  {journal} {\bibinfo  {journal} {J. Appl. Math Mech.}\ }\textbf
  {\bibinfo {volume} {9}},\ \bibinfo {pages} {49} (\bibinfo {year} {1929})},\
  \Eprint
  {https://arxiv.org/abs/https://onlinelibrary.wiley.com/doi/pdf/10.1002/zamm.19290090104}
  {https://onlinelibrary.wiley.com/doi/pdf/10.1002/zamm.19290090104}
  \BibitemShut {NoStop}%
\bibitem [{\citenamefont {Hill}(1952)}]{Hill_1952}%
  \BibitemOpen
  \bibfield  {author} {\bibinfo {author} {\bibfnamefont {R.}~\bibnamefont
  {Hill}},\ }\href {https://doi.org/10.1088/0370-1298/65/5/307} {\bibfield
  {journal} {\bibinfo  {journal} {Proc. Phys. Soc. Sect. A}\ }\textbf {\bibinfo
  {volume} {65}},\ \bibinfo {pages} {349} (\bibinfo {year} {1952})}\BibitemShut
  {NoStop}%
\bibitem [{\citenamefont {Hill}(1963)}]{HILL1963}%
  \BibitemOpen
  \bibfield  {author} {\bibinfo {author} {\bibfnamefont {R.}~\bibnamefont
  {Hill}},\ }\href
  {https://doi.org/https://doi.org/10.1016/0022-5096(63)90036-X} {\bibfield
  {journal} {\bibinfo  {journal} {J. Mech. Phys. Solids}\ }\textbf {\bibinfo
  {volume} {11}},\ \bibinfo {pages} {357 } (\bibinfo {year}
  {1963})}\BibitemShut {NoStop}%
\bibitem [{\citenamefont {Pugh}(1954)}]{pughratio}%
  \BibitemOpen
  \bibfield  {author} {\bibinfo {author} {\bibfnamefont {S.}~\bibnamefont
  {Pugh}},\ }\href {https://doi.org/10.1080/14786440808520496} {\bibfield
  {journal} {\bibinfo  {journal} {The London, Edinburgh, and Dublin
  Philosophical Magazine and Journal of Science}\ }\textbf {\bibinfo {volume}
  {45}},\ \bibinfo {pages} {823} (\bibinfo {year} {1954})},\ \Eprint
  {https://arxiv.org/abs/https://doi.org/10.1080/14786440808520496}
  {https://doi.org/10.1080/14786440808520496} \BibitemShut {NoStop}%
\bibitem [{\citenamefont {İyigör}\ and\ \citenamefont
  {Uğur}(2014)}]{pughcriticalelastic_Iyigor_BGValue}%
  \BibitemOpen
  \bibfield  {author} {\bibinfo {author} {\bibfnamefont {A.}~\bibnamefont
  {İyigör}}\ and\ \bibinfo {author} {\bibfnamefont {S.}~\bibnamefont
  {Uğur}},\ }\href {https://doi.org/10.1080/09500839.2014.970239} {\bibfield
  {journal} {\bibinfo  {journal} {Philos. Mag. Lett.}\ }\textbf {\bibinfo
  {volume} {94}},\ \bibinfo {pages} {708} (\bibinfo {year} {2014})},\ \Eprint
  {https://arxiv.org/abs/https://doi.org/10.1080/09500839.2014.970239}
  {https://doi.org/10.1080/09500839.2014.970239} \BibitemShut {NoStop}%
\bibitem [{\citenamefont {Wu}\ \emph {et~al.}(2019)\citenamefont {Wu},
  \citenamefont {Fecher}, \citenamefont {Shahab~Naghavi},\ and\ \citenamefont
  {Felser}}]{pughcriticalelastic_claudia}%
  \BibitemOpen
  \bibfield  {author} {\bibinfo {author} {\bibfnamefont {S.-C.}\ \bibnamefont
  {Wu}}, \bibinfo {author} {\bibfnamefont {G.~H.}\ \bibnamefont {Fecher}},
  \bibinfo {author} {\bibfnamefont {S.}~\bibnamefont {Shahab~Naghavi}},\ and\
  \bibinfo {author} {\bibfnamefont {C.}~\bibnamefont {Felser}},\ }\href
  {https://doi.org/10.1063/1.5054398} {\bibfield  {journal} {\bibinfo
  {journal} {J. Appl. Phys.}\ }\textbf {\bibinfo {volume} {125}},\ \bibinfo
  {pages} {082523} (\bibinfo {year} {2019})},\ \Eprint
  {https://arxiv.org/abs/https://doi.org/10.1063/1.5054398}
  {https://doi.org/10.1063/1.5054398} \BibitemShut {NoStop}%
\bibitem [{\citenamefont {Pettifor}(1992)}]{pettifor}%
  \BibitemOpen
  \bibfield  {author} {\bibinfo {author} {\bibfnamefont {D.~G.}\ \bibnamefont
  {Pettifor}},\ }\href {https://doi.org/10.1179/mst.1992.8.4.345} {\bibfield
  {journal} {\bibinfo  {journal} {Mater. Sci. Technol.}\ }\textbf {\bibinfo
  {volume} {8}},\ \bibinfo {pages} {345} (\bibinfo {year} {1992})},\ \Eprint
  {https://arxiv.org/abs/https://doi.org/10.1179/mst.1992.8.4.345}
  {https://doi.org/10.1179/mst.1992.8.4.345} \BibitemShut {NoStop}%
\bibitem [{\citenamefont {Hakimi}\ \emph {et~al.}(2010)\citenamefont {Hakimi},
  \citenamefont {Kameli},\ and\ \citenamefont
  {Salamati}}]{hakimi2010structural}%
  \BibitemOpen
  \bibfield  {author} {\bibinfo {author} {\bibfnamefont {M.}~\bibnamefont
  {Hakimi}}, \bibinfo {author} {\bibfnamefont {P.}~\bibnamefont {Kameli}},\
  and\ \bibinfo {author} {\bibfnamefont {H.}~\bibnamefont {Salamati}},\
  }\href@noop {} {\bibfield  {journal} {\bibinfo  {journal} {J. Magn. Magn.
  Mater.}\ }\textbf {\bibinfo {volume} {322}},\ \bibinfo {pages} {3443}
  (\bibinfo {year} {2010})}\BibitemShut {NoStop}%
\bibitem [{\citenamefont {Dubowik}\ \emph {et~al.}(2007)\citenamefont
  {Dubowik}, \citenamefont {Go{\'s}cia{\'n}ska}, \citenamefont {Kudryavtsev},\
  and\ \citenamefont {Oksenenko}}]{dubowik2007structure}%
  \BibitemOpen
  \bibfield  {author} {\bibinfo {author} {\bibfnamefont {J.}~\bibnamefont
  {Dubowik}}, \bibinfo {author} {\bibfnamefont {I.}~\bibnamefont
  {Go{\'s}cia{\'n}ska}}, \bibinfo {author} {\bibfnamefont {Y.}~\bibnamefont
  {Kudryavtsev}},\ and\ \bibinfo {author} {\bibfnamefont {V.}~\bibnamefont
  {Oksenenko}},\ }\href@noop {} {\bibfield  {journal} {\bibinfo  {journal}
  {Mater. Sci.}\ }\textbf {\bibinfo {volume} {25}} (\bibinfo {year}
  {2007})}\BibitemShut {NoStop}%
\bibitem [{\citenamefont {Dhakal}\ \emph {et~al.}(2020)\citenamefont {Dhakal},
  \citenamefont {Nepal}, \citenamefont {Ray}, \citenamefont {Paudel},\ and\
  \citenamefont {Kaphle}}]{our}%
  \BibitemOpen
  \bibfield  {author} {\bibinfo {author} {\bibfnamefont {R.}~\bibnamefont
  {Dhakal}}, \bibinfo {author} {\bibfnamefont {S.}~\bibnamefont {Nepal}},
  \bibinfo {author} {\bibfnamefont {R.}~\bibnamefont {Ray}}, \bibinfo {author}
  {\bibfnamefont {R.}~\bibnamefont {Paudel}},\ and\ \bibinfo {author}
  {\bibfnamefont {G.}~\bibnamefont {Kaphle}},\ }\href
  {https://doi.org/https://doi.org/10.1016/j.jmmm.2020.166588} {\bibfield
  {journal} {\bibinfo  {journal} {J. Magn. Magn. Mater.}\ }\textbf {\bibinfo
  {volume} {503}},\ \bibinfo {pages} {166588} (\bibinfo {year}
  {2020})}\BibitemShut {NoStop}%
\bibitem [{\citenamefont {Nepal}\ \emph {et~al.}(2020)\citenamefont {Nepal},
  \citenamefont {Dhakal},\ and\ \citenamefont {Galanakis}}]{our2}%
  \BibitemOpen
  \bibfield  {author} {\bibinfo {author} {\bibfnamefont {S.}~\bibnamefont
  {Nepal}}, \bibinfo {author} {\bibfnamefont {R.}~\bibnamefont {Dhakal}},\ and\
  \bibinfo {author} {\bibfnamefont {I.}~\bibnamefont {Galanakis}},\ }\href
  {https://doi.org/https://doi.org/10.1016/j.mtcomm.2020.101498} {\bibfield
  {journal} {\bibinfo  {journal} {Mater. Today Commun.}\ }\textbf {\bibinfo
  {volume} {25}},\ \bibinfo {pages} {101498} (\bibinfo {year}
  {2020})}\BibitemShut {NoStop}%
\end{thebibliography}%
\end{document}